\documentclass[12pt]{article}
\usepackage{mathrsfs}
\pdfoutput=1
\usepackage[english]{babel}
\newlength{\abstractwidth}
\usepackage{epsf}
\usepackage{color}
\usepackage{graphicx}
\usepackage{amssymb}
\usepackage{latexsym}

\usepackage{color}
\definecolor{dgreen}{rgb}{0,0.70,0.30}
\definecolor{gold}{rgb}{0.85,.66,0}
\definecolor{purple}{rgb}{1.0,0.3,0.6}
\definecolor{red}{rgb}{1.0,0.0,0.0}

\newcommand{\I}{i}

\newcommand{\pa}{\partial}

\newcommand{\nn}{\nonumber}

\newcommand{\eps}{\epsilon}
\newcommand{\IR}{\mathbb{R}}

\newcommand{\IZ}{\mathbb{Z}}

\def\ba{\begin{align}}
\def\ea{\end{align}}
\def\bse{\begin{subequations}}
\def\ese{\end{subequations}}
\def\Im{\,{\rm Im}\,}
\def\Re{\,{\rm Re}\,}

\def\cA{{\cal A}}
\def\cB{{\cal B}}

\def\cD{{\cal D}}
\def\cE{{\cal E}}
\def\cF{{\cal F}}

\def\cI{{\cal I}}

\def\cL{{\cal L}}
\def\cM{{\cal M}}

\def\cO{{\cal O}}

\def\cR{{\cal R}}
\def\cS{{\cal S}}

\def\cV{{\cal V}}

\def\bC{{\bf C}}

\def\ba{{\bf a}}

\def\mA{\mathfrak{A}}

\def\p{\partial}
\def\half{{1\over 2}}
\def\f{\varphi}
\def\no{\nonumber}
\def\sm{\smallskip}

\def\CC{{\mathbb C}}
 
\def\RR{{\mathbb R}}
\def\TT{{\mathbb T}}
\def\ZZ{{\mathbb Z}}

\def\mm{m_d}
\def\rr{\rho_d}
\def\gd{g_d}
\def\dd{\cD}

\def\tet{\vartheta}
\def\ep{\varepsilon}
\def\om{\omega}

\def\half{ {1\over 2}}

\renewcommand{\thefootnote}{\fnsymbol{footnote}}
\renewcommand{\thanks}[1]{\footnote{#1}}
\newcommand{\starttext}{
\setcounter{footnote}{0}
\renewcommand{\thefootnote}{\arabic{footnote}}}

\newcommand{\bea}{\begin{eqnarray}}
\newcommand{\eea}{\end{eqnarray}}
\newcommand{\be}{\begin{eqnarray}}
\newcommand{\ee}{\end{eqnarray}}

\newcommand{\tfrac}[2]{{\scriptsize \frac{#1}{#2}}}
\newcommand{\eqref}[1]{(\ref{#1})}
\begin{document}
\starttext
\setcounter{footnote}{0}

\begin{flushright}
2014 October 21  \\
NSF-KITP-14045\\
DAMTP-2014-22\\
QMUL-PH-14-11\\
CERN-PH-TH-2014-079\\
\end{flushright}

\bigskip

\begin{center}

{\Large \bf Matching the  $D^6 \cR^4$ interaction at two-loops}

\vskip .5in

{\large \bf Eric D'Hoker$^{(a)}$, Michael B. Green$^{(b)}$, Boris Pioline$^{(c)}$ \\ and  Rodolfo Russo$^{(d)}$}

\vskip .2in

{ \sl (a) Department of Physics and Astronomy }\\
{\sl University of California, Los Angeles, CA 90095, USA} \\
{\sl and Kavli Institute for Theoretical Physics}\\
{\sl University of California, Santa Barbara, CA 93106, USA}

\vskip 0.08in

{ \sl (b) Department of Applied Mathematics and Theoretical Physics }\\
{\sl Wilberforce Road, Cambridge CB3 0WA, UK}

\vskip 0.08in

{ \sl (c) Laboratoire de Physique Th\'eorique et Hautes
Energies, CNRS UMR 7589, \\ 
Universit\'e Pierre et Marie Curie,
4 place Jussieu, 75252 Paris cedex 05, France \\
and Theory Division, CERN, 
CH-1211, Geneva 23, Switzerland}

\vskip 0.08in

{ \sl (d)  Centre for Research in String Theory, School of Physics and Astronomy,\\
Queen Mary University of London, Mile End Road, London, E1 4NS, UK}

\vskip 0.1in

{\tt \small dhoker@physics.ucla.edu; M.B.Green@damtp.cam.ac.uk; boris.pioline@cern.ch; r.russo@qmul.ac.uk}

\end{center}

\vskip .2in

\begin{abstract}
 
The coefficient of the $D^6 \cR^4$  interaction in the low energy expansion 
of the two-loop four-graviton amplitude in  type II superstring theory is known 
to be proportional to the integral of the Zhang-Kawazumi  (ZK) invariant over 
the moduli space of genus-two Riemann surfaces. We demonstrate that the 
ZK invariant is an eigenfunction with eigenvalue $5$ of the Laplace-Beltrami 
operator in the interior of moduli space.  Exploiting this result, we evaluate the 
integral of the ZK invariant explicitly, finding agreement with the  value of the 
two-loop $D^6 \cR^4$ interaction predicted on the basis of S-duality and 
supersymmetry.  A review of the current understanding of the $D^{2p} \cR^4$ 
interactions in type II superstring theory compactified on a torus $T^d$
with $p\leq 3$ and $d \leq 4$ is included.

\end{abstract}

\newpage
\setcounter{tocdepth}{2}
\tableofcontents

\newpage

\section{Introduction}
\setcounter{equation}{0}
\label{intro}
  
Much of our current understanding of superstring theory is based on two different
 expansion schemes: the genus expansion, and the low-energy expansion.  
 Our aim is to compute a particular  term in this double expansion, namely the two-loop contribution 
 to the $D^6 \cR^4$ interaction in type II superstrings, 
 and check agreement with predictions from dualities and supersymmetry. 

\sm

The genus expansion, on the one hand,  is an asymptotic series in the string coupling, 
$g_s$, valid at small coupling.  A term of order $g_s^{2h-2}$ is associated with an integral 
over the moduli space of  genus-$h$ (super)Riemann surfaces,  generalizing  
$h$-loop Feynman diagrams in quantum field theory.  Obtaining explicit expressions 
for scattering amplitudes in string perturbation theory has  been possible up to genus-two
\cite{D'Hoker:2001nj, D'Hoker:2002gw} (see \cite{D'Hoker:2014nna} for a recent overview), 
but higher genus contributions remain largely elusive. Non-perturbative contributions 
to scattering amplitudes are in general unknown. 

\sm

The low-energy expansion, on the other hand, is  valid for small momenta, in units of the
inverse string length scale $1/\sqrt{\alpha'}$.   The leading term in this expansion reproduces the tree-level amplitudes of supergravity, while terms of higher order in $\alpha'$  describe local and non-local higher derivative effective  interactions. The coefficient in front of each of these terms  is a function of the moduli, including the string coupling $g_s$ and the parameters describing the target space, and receives both perturbative and non-perturbative contributions. The first few local interactions in the low-energy expansion are typically 
determined at low order in string perturbation theory, up to non-perturbative contributions 
which are constrained, and sometimes uniquely fixed, by supersymmetry and duality.

\sm

We will concentrate on  flat maximally supersymmetric backgrounds of type IIB  superstring theory.  
For the simplest such background, namely ten-dimensional Minkowski space-time, the amplitudes 
are   expected to be invariant under the action of the S-duality group $SL(2,\ZZ)$.  This group  
acts on the single complex  modulus field $T$ (the axion-dilaton)   by M\"obius  transformations. 
We shall also  consider partially compactified space-times of the form $\RR^{10-d} \times \TT^d$,
where $\TT^d$ is a flat torus of dimension $d$. The moduli space in this case includes,
in addition to the axion-dilaton $T$, the constant  metric $G$ and 2-form field $B$ on 
$\TT^d$, along with the Ramond--Ramond potentials  (and when $d\geq 6$, Neveu-Schwarz 
axions). The corresponding set of all moduli, denoted by $\mm$, parametrizes  
the symmetric coset spaces\footnote{\label{groupdef}The rank $n$ Lie group  $E_n(\RR)$  
coincides with the (non-compact) real split form $E_{n(n)}$ of the exceptional  Lie group $E_n$ 
for $n=8,7,6$, while  for the ranks $n=5,4,3,2,1$, the groups $E_n (\RR)$ are respectively given by $SO(5,5,\RR)$, $SL(5,\RR)$, 
$SL(3,\RR) \times SL(2,\RR)$, $SL(2,\RR) \times \RR^+$, and $ SL(2,\RR)$. The group 
$K_n (\RR)$ is the maximal compact subgroup of $E_n(\RR)$.} $E_{d+1}(\RR) / K_{d+1}(\RR)$.  
The  latter may be viewed as a fiber bundle over
$\IR^+ \times SO(d,d)/SO(d)\times SO(d)$, where the first factor corresponds to the 
$(10-d)$-dimensional  string coupling $g_d=g_s/\sqrt{\det G}$ while the second factor
 parametrizes the moduli $\rho_d = (G,B)$. 
For $d\ge 8$  the notion of moduli space becomes ill-defined.
 
\sm

Moreover, we restrict attention to higher derivative local interactions of the form  $D^{2p}\cR^4$, 
with $0\leq p \leq 3$, where  $\cR^4$ indicates the particular tensorial contraction of four 
Riemann tensors  dictated by supersymmetry, and $D^{2p}$ stands for a certain combination 
of  $2p$ covariant derivatives to be described below. 
Effective interactions with only two or three powers of the Riemann tensor are forbidden by 
supersymmetry in type II.  The restriction to $p \le 3$ ensures that the interactions are related by 
supersymmetry to fermionic vertices with strictly fewer than 32 fermions, and hence will be 
protected from non-BPS contributions. We will use the notation
$\cE_{(0,0)} (\mm)\, \cR^4$, $\cE_{(1,0)} (\mm)\, D^4 \cR^4$, and $\cE_{(0,1)} (\mm)\, D^6\cR^4$ to denote  these effective 
interactions\footnote{The linearised $D^2\cR^4$ contribution to the four-graviton amplitude 
vanishes for kinematic reasons and the non-linear interaction is believed to vanish identically.
The coefficients $\cE_{(m,n)} (m_d)$ multiply effective interactions schematically represented
by $D^{2p} \cR^4$ with $p=2m+3n$, as will be explained  in section \ref{sec2.1}. }  
in the Einstein frame. 

\sm

The  coefficients $\cE_{(m,n)} (m_d)$ are functions on the symmetric space   $E_{d+1}(\RR) / K_{d+1}(\RR)$, invariant under the dualities of string theory.  In particular, they are invariant under
the T-duality group $SO(d,d,\IZ)$, which leaves the  $(10-d)$-dimensional  string coupling
$g_d$ invariant. This symmetry therefore holds at each order in the expansion in powers
of $g_d$. A far more powerful statement is that the coefficients $\cE_{(m,n)}(m_d)$  
should also be invariant under an arithmetic subgroup $E_{d+1}(\ZZ)$ of $E_{d+1}(\RR)$ 
known as  U-duality \cite{Hull:1994ys,Obers:1998fb}. It arises by combining T-duality with the 
$SL(d+1,\IZ)$ symmetry manifest in the M-theory  lift of type II string theory, and reduces 
to the S-duality group $SL(2,\IZ)$ for $d=0$. U-duality relates  perturbative and non-perturbative 
contributions and,  along with constraints arising from supersymmetry,  often determines
the exact form of these coefficient functions. For example,
in ten-dimensional Minkowski space (namely $d=0$), $\cE_{(0,0)}(T)$ and $\cE_{(1,0)}(T)$,  are 
constrained by supersymmetry to be eigenmodes of the Laplacian on the upper half $T$-plane
with eigenvalues $3/4$ and $15/4$, respectively, and are therefore  proportional to the non-holomorphic Eisenstein series $E^\star(\frac{3}{2},T)$ and $E^\star(\frac{5}{2},T)$
\cite{ Green:1997tv, Pioline:1998mn, Green:1998by, Green:1999pu, Sinha:2002zr}. 
Similar results hold for $d>0$, with $\cE_{(0,0)}(m_d)$ and $\cE_{(1,0)}(m_d)$ being 
combinations of Eisenstein series of the U-duality group  \cite{Pioline:1998mn,Obers:1999um,Green:2010wi,Green:2010kv,Basu:2007ru,Basu:2007ck}. These Eisenstein series reproduce the known
perturbative contributions up to one and two-loop, respectively, and for $d=2$ can be
derived from one-loop and two-loop supergravity amplitudes in M-theory compactified
on $T^2$ \cite{Green:1997as,Green:1999pu}. Like any Eisenstein series, (or residue thereof) 
they are quasi-eigenmodes of the Laplace-Beltrami operator on $E_{d+1}(\RR) / K_{d+1}(\RR)$, see 
(\ref{eisenone})-(\ref{eisentwo}) below, although this property has not yet been derived from supersymmetry. 

\sm

The situation for the coefficient $\cE_{(0,1)}$ of the $D^6 \cR^4$ interaction is more challenging,
as it is clear from the analysis of 2-loop supergravity amplitudes that $\cE_{(0,1)}$ cannot be 
an eigenmode of the Laplacian, rather it must satisfy an inhomogeneous Laplace equation,
Eq. (\ref{eisenthree}) below, with 
a source proportional to the square of  $\cE_{(0,0)}$, see \cite{Green:2005ba,Green:2010wi,Green:2010kv}. 
The solution of this equation contains perturbative terms that correspond to zero-, one-, two-, and  three-loop contributions in superstring theory.  The tree-level  and one-loop values were verified by perturbative string theory calculations for $d=0$ in \cite{Green:1999pv}, and further agreement has also been found for $d\le 4$  \cite{Green:2010wi} and $d=7$ \cite{Green:2010kv}. Furthermore, the predicted three-loop value agrees with the value of the  coefficient  of  $D^6\cR^4$ in the type IIA theory predicted by an S-duality argument  in \cite{Green:1999pu}.  There is also a  claimed agreement with an explicit  three-loop string theory calculation in \cite{Gomez:2013sla}  (although a potential mismatch by a factor of 3 needs to be sorted out).  

\sm

The only remaining S-duality prediction still to be verified in superstring perturbation theory 
is the two-loop contribution to the coefficient of $D^6\cR^4$, which will be denoted by
$\cE_{(0,1)}^{(2)}$.  The predicted value for $d=0$ is,   
\be
\cE_{(0,1)}^{(2)}=\frac{2\pi^4}{45}
\label{predicttwo}
\ee

\sm

This value should emerge from the low energy expansion of the two-loop four-graviton string 
amplitude obtained in \cite{D'Hoker:2001nj, D'Hoker:2005jc}\footnote{The construction of the two-loop
gauge invariant measure was obtained in \cite{D'Hoker:2001zp, D'Hoker:2001qp}; a construction
of the measure based on holomorphy and modular invariance was given in \cite{Witten:2013tpa};
gauge invariance of the two-loop four-graviton amplitude was proven in  \cite{D'Hoker:2005jb}; 
the amplitude was reproduced in the pure spinor formulation and extended to including external 
fermions in \cite{Berkovits:2005ng}}.  While the two-loop 
contributions to  the $\cR^4$ and $D^4\cR^4$ interactions were relatively straightforward to 
extract from the explicit two-loop four-graviton amplitude \cite{D'Hoker:2005ht},  the two-loop 
contribution  to $D^6\cR^4$ involves various mathematical subtleties. 
As a step towards its evaluation, the value of $\cE_{(0,1)}^{(2)}$ arising from an analysis of the 
two-loop amplitude  was expressed  in \cite{D'Hoker:2013eea} as the integral,
\bea
\label{1b}
\cE_{(0,1)} ^{(2)} = \pi \int _{\cM_2} d \mu_2 \, \varphi
\eea
Here  $d \mu_2$  is the
canonical volume form on $\cM_2$ and $\f$ may be expressed as follows,
\bea
\label{defphi}
\f (\Sigma)  = - { 1 \over 8} \int _{\Sigma ^2} P(x,y) \, G(x,y)
\eea
where $G(x,y)$ is the scalar Green function on the genus-two surface, $\Sigma$, 
and $P(x,y)$ is a bi-holomorphic form in $x,y \in \Sigma$ that is defined  in a manner 
that makes (\ref{defphi}) conformal invariant, as  will be reviewed in section \ref{zkprops}. 
The object $\f$ in (\ref{defphi}) is an invariant of the Riemann surface\footnote{The definition 
(\ref{defphi}) holds for a Riemann surface of any genus $h\geq 2$, upon replacing $8$ by 
$4h$ in the denominator. The ZK invariant vanishes for $h=0,1$.} $\Sigma$
that  was discovered  by Zhang  \cite{Zhang} and Kawazumi \cite{Kawazumi}.
The equivalence of (\ref{defphi}) with other definitions of the Zhang--Kawazumi (ZK)  invariant in the mathematical literature was derived in \cite{D'Hoker:2013eea}. In spite a variety of
available reformulations, the direct integration of $\varphi$ on the moduli space $\cM_2$ of 
genus-two Riemann surfaces remained elusive.

\sm

In this paper, we shall evaluate the integral in (\ref{1b}) and thereby find agreement between the 
value predicted by S-duality, (\ref{predicttwo}), and the result of  two-loop superstring 
perturbation theory.  This matching is highly non-trivial, and involves some novel mathematics. 
The key observation which makes this computation possible is the fact that $\varphi$ satisfies 
the following differential equation everywhere in the interior of $\cM_2$,
\bea
\label{1c}
(\Delta - 5) \, \f=0
\eea
where $\Delta $  is the Laplace-Beltrami operator on $\cM_2$ associated with the Poincar\'e
metric on the Siegel upper half space $\cS_2$, to be defined in Appendix \ref{appA} below. 
Equation (\ref{1c}) will be derived here from 
first principles by using the theory of deformation of complex structures. The validity of (\ref{1c}) 
may be extended to the Deligne-Mumford compactification $\overline{\cM_2}$  of  moduli space, 
upon supplementing (\ref{1c}) by a term with $\delta$-function support on the separating node,
\bea
\label{1d}
(\Delta - 5) \, \f= - 2 \pi \delta _{SN}
\eea
an equation which is valid throughout $\overline{\cM_2}$. We note that the combination
$ \partial \bar\partial  \f$ was evaluated in \cite{Kawazumi,DeJong2} for arbitrary genus.
It would be interesting to understand its  connection with  (\ref{1c}) and (\ref{1d}), if any exists,
especially since the  ZK invariant for genus $h\geq 3$ does not satisfy a simple 
equation of the type (\ref{1c}), as will be shown in Appendix C.

\subsection{Outline}

The layout of this paper is as follows.   In section \ref{overview} we give a brief overview 
of perturbative and non-perturbative  features of the low energy expansion of the four-graviton 
amplitude, both in flat 10-dimensional space-time and upon compactification on $d$-dimensional 
flat tori with $d \leq 4$. The differential equations imposed by S-duality and supersymmetry 
on the coefficient 
functions, $\cE_{(m,n)}(\mm)$, are reviewed, and their implications on the perturbative 
expansion of these coefficients are obtained. In section \ref{zkprops}, the differential constraints 
mentioned above are recast in the form of constraints on certain integrals of the  ZK invariant $\f$,
which suggest a Laplace eigenvalue equation for $\f$. Further evidence is 
provided by the fact that the supergravity limit of $\f$ satisfies this equation.
In section \ref{laplacezk}, the Laplace eigenvalue equation for $\f$ is proven 
for genus-two using the methods of deformations of complex structures on Riemann surfaces. 
In section \ref{intvarphi} the Laplace eigenvalue equation is used to integrate $\f$
over moduli space, and thereby prove the matching of the coefficient of the $D^6\cR^4$
interaction. We end by providing the corresponding differential equations  for 
other genus-two modular forms including the Faltings invariant.
Basic facts about modular geometry are collected in Appendix A;   details of the calculation of the Laplacian 
are given in Appendix B while its generalization to higher genus is obtained in Appendix C.

\newpage

\section{Constraints from S-duality and supersymmetry} 
\setcounter{equation}{0}
\label{overview}

In this section, we shall  describe relevant aspects of the $\alpha '$ expansion for the four-graviton
amplitude in type IIB string theory compactified on $\RR^{10-d} \times \TT^d$ for $0 \leq d \leq 4$, 
and of the differential equations satisfied by the coefficient 
functions $\cE_{(m,n)}$ of the various BPS effective interactions.

\subsection{The full four-graviton amplitude}
\label{sec2.1}

The full four-graviton amplitude $\cA ^{(4)} (\eps_i, k_i ; \mm)$ depends on the $(10-d)$-dimensional  
polarization tensors $\eps_i $ and momenta $k_i$ for $i=1,2,3,4$, as well as on the moduli $\mm$. 
The dependence on $d$ will be understood throughout, but not exhibited. The maximal 
supersymmetry of type IIB string theory singles out a unique tensorial structure for 
the full four-graviton amplitude which we denote by $\cR^4$, and which is a special 
contraction of four Riemann tensors. Its explicit form may be found in \cite{D'Hoker:2013eea},
for example, but will not be needed here. As a result, we may introduce a reduced 
four-graviton amplitude $\cI$, defined by,
\bea
\label{A4full}
\cA ^{(4)} (\eps_i, k_i ; \mm) = \kappa_d ^2 \, \cR^4 \, \cI (s,t,u;\mm)
\eea
where $\kappa_d^2$ is Newton's constant in $10-d$ dimensions.  The  reduced 
amplitude $\cI$ is a dimensionless function of the moduli $\mm$ and the dimensionless variables 
$s=-\alpha ' (k_1+k_2)^2/4$, $t=-\alpha ' (k_1+k_4)^2/4$ and $u = -\alpha ' (k_1+k_3)^2/4$,
subject to the relation $s+t+u=0$ which follows from momentum conservation
and the mass-shell conditions $k_i^2=0$.
Our interest is in the low energy expansion of $\cI$, or $\alpha'$ 
expansion,  in which $|s|, |t| , |u| \ll 1$.

\sm

In this limit, the reduced amplitude  $\cI (s,t,u; \mm)$ can be decomposed into a sum of terms that are analytic  in $s,t,u$, corresponding to local interactions in the effective action, and terms that are non-analytic, corresponding to non-local effective interactions. Non-local terms arise from integrating out massless states, and are computable in the framework of  pertubative supergravity, supplemented by appropriate counter-terms to remove ultraviolet divergences. The remainder is an analytic function of $s,t,u$ which contains contributions from
massive string states. The local part, of interest in this work, is related to, and constrained by, the non-local part via unitarity. In the Einstein frame, the local part has an expansion of the form,
\bea
\cI (s,t,u; \mm) \bigg | _{\rm local}=  \frac{3}{\sigma_3} +
\sum _{m,n=0}^\infty   \cE_{(m,n)} (\mm) \, \sigma _2^m \, \sigma _3^n 
\label{genexpan}
\eea 
The first term in (\ref{genexpan}) gives the tree-level amplitude of   type II supergravity, 
while the higher-order terms give higher-derivative local interactions.  
Because of the relation $s+t+u=0$, the expansion is in powers of only two independent 
variables, chosen to be $\sigma_2=s^2+t^2+u^2$ and $\sigma_3=s^3+t^3+u^3$ \cite{Green:1999pv}.
The variable $\gd$ is the effective string coupling defined by $\gd=g_s$ when $d=0$ and 
$\gd= g_s/\sqrt{\cV_d}$ for $d>0$ where $\cV_d$ is the volume of $\TT^d$ with metric $G$. 

\sm

In the Einstein frame, the reduced amplitude $\cI (s,t,u; \mm)$ and the dimensionless  
variables $s,t,u$ are  invariant under U-duality. Therefore, the coefficients $\cE_{(m,n)}(\mm)$ 
must be automorphic forms on  $E_{d+1}(\ZZ)\backslash E_{d+1}(\RR)/K_{d+1}(\RR)$. The effective interactions produced by the lowest order terms  
$\cE_{(0,0)} (\mm) \cR^4$, $\cE_{(1,0)} (\mm) D^4 \cR^4$, and $\cE_{(0,1)} (\mm) D^6 \cR^4$
will be of central interest in this paper.

\subsection{Perturbative contributions to the low energy expansion}
\label{sec:pert}

Features of string perturbation theory, which are the concern of this paper, 
are obtained by expanding the coefficient functions $\cE_{(m,n)}(\mm)$ in powers of $\gd$.  
Each term in this expansion is then a function on the coset
 \be
 \label{rhocoset}
 \rr =G+B \in SO(d,d,\RR)/SO(d,\RR)^2
 \ee
parametrized by the metric $G$ and two-form field $B$ on $\TT^d$, automorphic under
 the T-duality group $SO(d,d,\ZZ)$. Mathematically, these perturbative contributions are 
 obtained   by  extracting the constant 
 term of the automorphic function $\cE_{(m,n)}$ with respect to the maximal parabolic
 sub-group $P_d$ of $E_{d+1}(\RR)$ whose Levi sub-group is $L_d=SO(d,d,\RR) \times \RR^+$,
 where  the factor $\RR^+$ corresponds to the coupling $\gd$. 
 In general however,
 this constant term may include non-analytic terms in $\gd$, such as powers of $\log\gd$.
 These terms do not exist in string perturbation theory, which is formulated by definition in the string frame, but may arise when rescaling the amplitude to the Einstein frame, due to mixing between
 the local and non-local parts of the low-energy effective action \cite{Kiritsis:1997em,Green:2010sp}. 

\sm

Finally, there are also exponentially suppressed contributions of order $\cO (e^{-2\pi/\gd})$ 
or smaller, with non-trivial dependence on the  Ramond-Ramond moduli. These contributions 
are interpreted as D-brane instanton corrections (along with NS-brane instantons when $d\geq 6$). Mathematically, they correspond to non-zero Fourier coefficients with respect to the unipotent 
radical $P_d$.

\sm

As a result, the structure of the perturbative expansion of the coefficients 
$\cE_{(m,n)}(\mm)$ of the higher-derivative local interactions takes the following general form,
\bea
\label{2k4}
\cE_{(m,n)}(\mm) =  \gd ^{-\nu} 
\sum _{h=0} ^\infty \gd^{-2+2h} \, \cE_{(m,n)} ^{(h)} (\rr)
+ \cE_{(m,n)}^{\rm non-an.}(g_d,\rr)
+ \cO (e^{-2\pi/\gd})
\eea
where the contribution $\cE_{(m,n)} ^{(h)} (\rr)$ arises to $h$-loop order in superstring 
perturbation theory.   The overall factor $\gd^{-\nu} $, with $\nu = (2d-4+8m+12n)/(8-d)$, 
converts the genus $h$ contribution from the string frame to the $(10-d)$-dimensional 
Einstein frame. Our main focus in this work is on the perturbative contributions 
$\cE_{(m,n)} ^{(h)}$, but it is important to take into account the non-analytic contribution 
(when present) as it affects the differential equations satisfied by the functions $\cE_{(m,n)} ^{(h)}$.

\sm

The four-graviton scattering amplitude in superstring perturbation theory at arbitrary $h$-loop 
order involves an integral over the moduli space of super-Riemann surfaces of genus $h$ 
with four punctures. In favorable cases (which include $h=1$ and $h=2$), this integral 
may be reduced to an integral over the moduli space $\cM_h$ of ordinary compact Riemann surfaces
of genus $h$, and parametrized by the period matrix $\Omega$ of the surface $\Sigma$, 
which takes values in the Siegel upper-half plane $\cS_h$  (subject, for $h>3$,
to Schottky relations).  For the interactions $\cE_{(m,n)}$ of interest in this paper, 
the interplay between supersymmetry and dualities implies that the perturbative 
coefficients  are non-zero for only a finite number of loop orders $h$, namely,
\bea
\cE _{(0,0)} ^{(h)} (\rr) =0 & \hskip 1in & h \geq 2
\no \\
\cE _{(1,0)} ^{(h)} (\rr) =0 & \hskip 1in & h \geq 3
\no \\
\cE _{(0,1)} ^{(h)} (\rr) =0 & \hskip 1in & h \geq 4
\eea
Our conventions for  the integration measure  $d \mu_h$ on $\cM_h$ are described in detail in 
Appendix A. It is customary to express  $\Omega$ in terms of real 
matrices $X,Y$ defined by $ \Omega = X + i Y$, and we shall do so also here throughout.
\sm

A key ingredient in genus $h$ amplitudes in superstring theory compactification on a torus $\TT^d$  is the partition function 
for the zero-modes of the compact bosons on a genus $h$ worldsheet,  
$\Gamma _{d,d,h} (\rho_d;\Omega)$, given  by the following standard lattice sum,
 \be
\label{Gamddh}
\Gamma_{d,d,h}  (\rr ;\Omega) =(\det Y)^{d/2} \!\!\!\!\!
 \sum_{m_\alpha ^I,n^{\alpha I} \in \IZ}
   \!\!\!\!\!
 \exp \Big \{ -\pi \cL^{IJ}(\rr)  Y_{IJ}+2\pi\I m_\alpha ^I n^{\alpha J} X_{IJ} \Big\}
\ee
where the quadratic form in $m,n$ is defined by, 
\be
\cL^{IJ} (\rr) =(m_\alpha ^I+B_{\alpha \gamma } n^{\gamma I}) 
G^{\alpha \beta }  (m_\beta ^J+B_{\beta \delta } n^{\delta J}) + n^{\alpha  I} G_{\alpha \beta}  n^{\beta J} 
\ee
The integers $m_\alpha ^I$ and $n^{\alpha I}$ label momenta and windings.
The range of the indices is $I,J=1,\cdots, h$ and $\alpha, \beta, \gamma, \delta = 1, \cdots, d$,  
and repeated indices  are to be summed over. Note that we have $\Gamma _{0,0,h}=1$.

\sm

In order to set the scene for the body of this paper, we list  the results of explicit string perturbation theory calculations (a somewhat more detailed review is contained in  \cite{D'Hoker:2013eea}).

\subsubsection{Genus zero}

The tree-level amplitude can easily be expanded to all orders in the low energy expansion 
and the coefficients are independent of the moduli.  Normalizing the classical 
tree-level term as in (\ref{genexpan}),  the subsequent tree-level 
coefficients are given in terms of the Riemann zeta function $\zeta (s)$ evaluated at odd integers by,
\be
\label{treecoeffs}
\cE_{(0,0)} ^{(0)} (\rr) = 2 \zeta(3) \hskip 0.6in 
\cE_{(1,0)} ^{(0)} (\rr) =  \zeta(5) \hskip 0.6in
\cE_{(0,1)} ^{(0)} (\rr) = \frac{2}{3} \zeta(3)^2
\ee
No dependence on the moduli $\rr$ arises at tree-level since the momenta and polarization tensors 
of the four gravitons are along the subspace $\RR^{(10-d)}$, but not along $\TT^d$.

\subsubsection{Genus one}

The one-loop amplitude is an integral over the complex structure of the world-sheet torus 
and the contributions of terms in the low energy expansion  reduce to integrals over the 
fundamental domain $\cM_1$ of the Poincar\'e upper half plane \cite{Green:1999pv,Green:2008uj}, 
given by,
\bea
\label{rfournew}
\cE_{(0,0)} ^{(1)} (\rr) & =& \pi\, \int_{\cM_1} d \mu_1\, \Gamma_{d,d,1} (\rr;\tau)  
\\
\label{dfourrfournew}
\cE_{(1,0)} ^{(1)} (\rr) & =& 
2\pi \, \int_{\cM_1} d \mu_1 \, \Gamma_{d,d,1}(\rr;\tau) \, E^\star(2,\tau)
\\
\label{dsixrfournew}
\cE_{(0,1)} ^{(1)} (\rr)  &=& 
\frac{\pi}{3}\, \int_{\cM_1} d \mu_1 \, \Gamma_{d,d,1}(\rr;\tau) \, \left( 5
E^\star(3,\tau) +\zeta(3) \right)
\eea
The  modulus $\tau$ parametrizes the genus-one moduli space $\cM_1$,
and its volume form $d \mu_1$, both of which are given in Appendix A.
The factor $\Gamma_{d,d,1}(\rr;\tau)$ is the genus-one partition function on $\TT^d$
defined in (\ref{Gamddh}) for general $h$. The quantity,
 \be
\label{eisndef}
E^\star(s,\tau)=\frac12 \pi ^{-s} \Gamma (s) \zeta (2s)\sum_{(c,d)=1} \frac{(\Im \tau)^s}{|c\tau+d|^{2s}}
\ee
is a non-holomorphic Eisenstein series, in the normalization of 
\cite{Angelantonj:2011br}.   

\sm

The integrals (\ref{rfournew})-(\ref{dsixrfournew}) are not generally convergent due to the 
polynomial growth of the integrand as $\Im \tau \to \infty$.  Specifically, the  integral in (\ref{rfournew})  
diverges for $d\geq 2$, while the integrals in (\ref{dfourrfournew}) and  (\ref{dsixrfournew})  
likewise diverge when $d \geq 0$ since $E^\star(s,\tau) = \cO((\Im \tau)^{{\rm max} (s,1-s)})$.  
These divergences reflect the presence of a non-local term of the form $s^{(D-8)/2} \cR^4$,
(times $\log s$ in even dimension $D\geq 8$),  produced by a one-loop
infrared threshold, which dominates over the local term when $D\leq 8$.
The prescription  used in \cite{Green:1999pv,Green:2008uj} to separate the local 
and non-local contributions  leads to a particular renormalization prescription for these integrals, 
which is equivalent to one used in the Rankin-Selberg-Zagier method  \cite{Zagier1982,Angelantonj:2011br}. In particular, for $d=0$, the renormalized integral
 $\int_{\cM_1} d \mu_1 \, E^\star(s,\tau)$ vanishes for any $s$, so that we have,
\be
\label{oneloopcoeffs}
\cE_{(0,0)} ^{(1)}  =  4 \zeta(2) \hskip 0.6in 
\cE_{(1,0)} ^{(1)}  =  0 \hskip 0.6in
\cE_{(0,1)} ^{(1)}  = { 4 \over 3} \zeta (2) \zeta(3)
\ee
We have suppressed the dependence on $\rr$ since there are no such moduli for $d=0$.
For $0< d \leq 4$, the result can instead be expressed in terms of the Eisenstein series 
$E^{SO(d,d)}_{V,s} (\rr) $ associated with the parabolic subgroup of $SO(d,d)$ that is 
labelled by the weight $V$ of the vector representation of $SO(d,d)$,
\bea
\label{E001gen}
\cE_{(0,0)}^{(1)} (\rr) &=& 2\pi^{2-\tfrac{d}{2}} \Gamma \left ( \tfrac{d}{2}-1 \right )\, 
E^{SO(d,d)}_{V,\tfrac{d}{2}-1} (\rr) \qquad (d\neq 2)
\\
\label{E101gen}
\cE_{(1,0)}^{(1)} (\rr) &=& \frac{2}{45} \pi^{2-\tfrac{d}{2}} \Gamma \left ( 1+\tfrac{d}{2} \right )\, 
E^{SO(d,d)}_{V,\tfrac{d}{2}+1}(\rr)\qquad\qquad (d\neq 4)
\\
\label{E011gen}
\cE_{(0,1)}^{(1)} (\rr) &=& \frac{\zeta(3)}{3}\, \cE_{(0,0)}^{(1)} (\rr)+
\frac{4}{567} \pi^{2-\tfrac{d}{2}} \Gamma \left ( \tfrac{d}{2}+2 \right )\, 
E^{SO(d,d)}_{V,\tfrac{d}{2}+2}  (\rr)
\eea
The excluded values of $d$ are those for which  the integral is logarithmically divergent 
and the Eisenstein series has a pole. In that case, the formulae 
(\ref{E001gen})-(\ref{E011gen}) hold after subtracting the pole.

\subsubsection{Genus two}

Since this is the case of central interest in this paper we will review it in somewhat more detail.
 The full two-loop four-graviton amplitude is given by \cite{D'Hoker:2005ht},  
 \bea
\label{gen2amp}
\cA_2 ^{(4)} (\epsilon_r, k_r; \gd, \rr)
= 
{ \pi \over 64} \kappa ^2_d  \, g_s^2  \, \cR^4 
\int_{\cM_2} d\mu_2 \, \cB_2 (s,t,u ; \Omega)\, \Gamma_{d,d,2}(\rr;\Omega)
\eea
The reduced amplitude $\cB_2$   is given by,
\bea
 \label{2a55}
\cB_2  (s,t,u ; \Omega)
 =   \int_{\Sigma^4} { |{\cal Y}_S|^2 \over (\det Y )^2}
\exp \left  \{ -\frac{\alpha'}{2} \sum_{i<j} k_i \cdot k_j\,G(z_i,z_j) \right \} 
\eea
The integration is over four copies $\Sigma ^4$ of the genus-two Riemann 
surface $\Sigma$.  The quantity $\cal Y_S$ in the measure in (\ref{2a55}) is a $s,\, t, \, u$-dependent 
family of  holomorphic sections of the canonical line bundle $K$ over $\Sigma$ in each vertex 
insertion point $z_i$  for $i=1,2,3,4$,  as defined in \cite{D'Hoker:2005jc}. 
The lattice partition function $\Gamma _{d,d,2}$ was defined in (\ref{Gamddh}). 

\sm

The lowest order genus-two contribution $\cE_{(0,0)} ^{(2)} (\rr) $ corresponds to the 
effective interaction $\cR^4$ and vanishes in any dimension \cite{D'Hoker:2005jc},
\be
\cE_{(0,0)} ^{(2)} (\rho_d) =  0 
\ee 
The next contribution corresponds to 
$D^4 \cR^4$  and is given by  \cite{D'Hoker:2005ht}, 
\be
\cE_{(1,0)} ^{(2)} (\rr)  = \frac{\pi}{2} \, \int_{\cM_2}  d \mu_2 \, \Gamma_{d,d,2} ( \rr; \Omega)
\ee
The integral is infrared divergent for $d\geq 3$, due to the
presence of a two-loop non-local term of the form $s^{D-7} D^4 \cR^4$,
(times $\log s$ in $D=7$),  which dominates over the local term when $D\leq 7$.
The renormalized 
integral can in principle be defined by a genus-two version of the Rankin-Selberg-Zagier 
prescription. The result may be expressed as the residue  of the Langlands-Eisenstein series 
associated with the weight of the two-index antisymmetric representation of $SO(d,d)$ at $s=d/2$ \cite{Pioline:2014bra}. Alternatively, the conjectured results of \cite{Obers:1999um}, further supported in  \cite{Green:2010wi,Pioline:2014bra}, may be used to express the result as,
\be
\label{E102gen}
\cE_{(1,0)}^{(2)} (\rr)= \frac23 \left( \hat E^{SO(d,d)}_{S,2} (\rr) + \hat E^{SO(d,d)}_{C,2} (\rr) \right)
\ee
where $E^{SO(d,d)}_{S,s}(\rr) $ and $E^{SO(d,d)}_{C,s}(\rr) $ are the Eisentein series associated 
with the two distinct spinor weights $S$ and $C$ of $SO(d,d)$, and the hat
indicates that the simple pole at $s=2$ has been subtracted.  For $d=0$, one has, 
\be
\cE_{(1,0)} ^{(2)}  =  {4 \over 3} \zeta (4)
\ee
in agreement with supersymmetry and S-duality, upon using the values of $\zeta $ 
given in (\ref{zeta}). 

\sm

Our primary interest in this paper  is the next term in the low energy expansion. It is  the 
genus-two contribution to $D^6 \cR^4$ that  was recently shown to have the form  
\cite{D'Hoker:2013eea},
\be
\cE_{(0,1)} ^{(2)} (\rr)  
= \pi\, \int_{\cM_2} d \mu_2 \, \Gamma_{d,d,2}(\rr; \Omega) \, \varphi(\Omega)
\label{dsixgentwo}
\ee
where $\varphi(\Omega)$ is the ZK invariant, whose form will be reviewed below.
The integral over $\cM_2$ is convergent for $d<2$, but has
both primitive and one-loop subdivergences in $d\geq 2$. This is 
consistent with  the presence of non-local terms of the form $s^{(D-8)/2} 
D^6 \cR^4$ and $s^{D-8} D^6 R^4$ (times $\log s$ when the power of $s$ is 
integer). These contributions dominate over the local terms when $D\leq 
8$. For $d \geq 2$, the integral must be renormalized, and it is not
known at present how to express it in terms of Eisenstein series of $SO(d,d,\IZ)$.
For $d=0$ and $d=1$, however,  the values predicted by S-duality and supersymmetry are as follows \cite{Green:2005ba,Green:2010wi}
\bea
\label{twoloopcoeffs}
\cE_{(0,1)} ^{(2)}  &=& { 8 \over 5} \zeta (2)^2  \qquad (d=0) \\
\cE_{(0,1)} ^{(2)}  (\rr) &=& { 8 \over 5} \zeta (2)^2 \left( r^2 + \frac{1}{r^2} + \frac{5}{3} \right) \qquad (d=1) 
\eea
where $r$ is the radius of the $d=1$ circle in string units.
The evaluation of the integral (\ref{dsixgentwo}) for the case $d=0$ is the main focus 
of the subsequent sections of this paper.

\subsubsection{Genus three}

The three-loop contribution to $D^6 \cR^4$ in ten-dimensional type IIB
string theory was recently computed in the pure spinor formalism
\cite{Gomez:2013sla}, and claimed to take the value for $d=0$,
\be
\cE_{(0,1)} ^{(3)} = \frac{4}{27}\, \zeta(6)
\label{genusthreecoeff}
\ee
in  agreement with the predictions from S-duality and supersymmetry
\cite{Green:2005ba}.  However, at present there are unresolved issues concerning 
a factor of three in the derivation of this value.  Assuming the value is indeed correct, 
a straightforward generalization to the theory compactified on a torus $\TT^d$ leads to,
\be
\label{D6R43}
\cE_{(0,1)} ^{(3)} (\rr)  = \frac{5}{16}\, \int_{\cM_3}  d \mu_3 \,
\Gamma_{d,d,3}(\rr; \Omega)
\ee
where $\cM_3$ is the fundamental domain of genus-three Riemann surfaces, and
$\Gamma_{d,d,3}(\rr; \Omega)$ is the genus 3 lattice sum (\ref{Gamddh}). Using the volume of
$\cM_3$ stated in \eqref{vols}, it is easily seen that \eqref{D6R43}  reduces to  
(\ref{genusthreecoeff}) when $d=0$.  In general, the modular integral in
(\ref{D6R43}) can be computed by a genus 3 extension of the Rankin-Selberg-Zagier 
method, and expressed as a residue of the Langlands-Eisenstein series
associated with the weight of the three-index antisymmetric tensor representation of
$SO(d,d)$ at $s=d/2$ \cite{Pioline:2014bra}. Alternatively, using the  
result conjectured in \cite{Obers:1999um} and further supported in  \cite{Green:2010wi}, one has,
\be
\label{E013gen}
\cE_{(0,1)}^{(3)}(\rr)  = \frac2{27} \left( \hat E^{SO(d,d)}_{S,3} (\rr) + \hat E^{SO(d,d)}_{C,3} (\rr) \right)
\ee
and the hat indicates that the simple pole at $s=3$ has been subtracted.

\subsection{S-duality and differential constraints}
\label{sec:sdual}
  
The three leading effective interactions in the $\alpha'$ expansion of (\ref{genexpan}),
namely $\cR^4$, $D^4 \cR^4$, and $D^6 \cR^4$,  are BPS-saturated interactions.  
The  exact coefficients  $\cE_{(0,0)}(\mm)$ and $\cE_{(1,0)}(\mm)$ of the $\cR^4$ 
and $D^4\cR^4$, including all non-perturbative corrections, have been conjectured 
based on variety of arguments including duality invariance, perturbative and  
decompactification  limits,  unitarity, and supersymmetry constraints.   

\sm

For example,  in ten-dimensional type IIB string theory ($d=0)$ the coefficient  
$\cE_{(0,0)}(T)$  is proportional to the non-holomorphic $SL(2,\IZ)$ Eisenstein 
series $E^\star(\frac32,T)$ \cite{Green:1997tv}. It is the unique solution of the 
Laplace equation $(\Delta_{SL(2)}-\frac34)\cE_{(0,0)}(T)=0$ on the upper half $T$ plane,  
a constraint which follows from a careful implementation of nonlinear extended 
supersymmetry \cite{Pioline:1998mn,Green:1998by}.  This also extends to 
$\cE_{(1,0)}(T)$ in $d=0$,  which is proportional to the Eisenstein series 
$E^\star(\frac52,T)$ \cite{Green:1998by,Sinha:2002zr}.   More generally, 
the conjectured coefficients $\cE_{(0,0)}(\mm)$ and $\cE_{(1,0)}(\mm)$ 
are given by linear combinations of Eisenstein series (and derivatives thereof) under the duality group 
$E_{d+1}(\ZZ)$. They  satisfy Laplace eigenvalue equations 
 \cite{Green:1997tv,Green:1999pu,Pioline:1998mn, Obers:1999um,Green:2005ba, Green:2010wi},
\bea
\label{eisenone}
\left ( \Delta_{E_{d+1}} - \frac{3(d+1)(2-d)}{(8-d)} \right )\,
\cE_{(0,0)} (\mm) & = & 6\pi\, \delta_{d,2}
\\
\left ( \Delta_{E_{d+1}} - \frac{5(d+2)(3-d)}{(8-d)} \right )\,
\cE_{(1,0)} (\mm) & = & 40 \, \zeta(2)\, \delta_{d,3} + 7\,  \cE_{(0,0)}\, \delta_{d,4}
\label{eisentwo}
\eea
where\footnote{
Our normalization  for the Laplace-Beltrami operators 
$\Delta_{E_{d+1}}$ and $\Delta_{SO(d,d)}$  differ by a factor of 2 from the ones used in  
\cite{Obers:1999um,Angelantonj:2011br,Pioline:2014bra,Pioline:2010kb}.}
 $\Delta_{E_{d+1}}$ is the  Laplace-Beltrami operator on the moduli space 
 $E_{d+1}(\RR) /K_{d+1}(\RR)$. It is expected that these equations
are consequences of non-linear supersymmetry, although this has not been fully established yet. 
The anomalous terms on the right-hand sides occur in dimensions in which the eigenvalues 
vanish, or when the eigenvalue of $\cE_{(1,0)}$ becomes degenerate with that of $\cE_{(0,0)}$.  
These terms  are correlated with logarithmic infrared divergences in string theory 
and in supergravity. Consequently, they are also correlated with the onset of ultraviolet divergences
in supergravity \cite{Green:2010sp}. 

\sm

For the $D^6 \cR^4$ interaction, a candidate for the exact coefficient $\cE_{(0,1)}(\mm)$ 
is only available in dimensions $d=0,1,2$, and in a rather implicit form 
\cite{Green:2005ba, Green:2010wi,Green:2014yxa}. General arguments, supergravity limits, 
and many consistency checks suggest that $\cE_{(0,1)}(\mm)$ is a solution of  
the inhomogeneous Laplace eigenvalue equation,\footnote{The $\delta_{d,4}$ terms in 
 (\ref{eisentwo}) and (\ref{eisenthree}) have been corrected from the ones given in equations (2.8) and (2.9) in \cite{Green:2010sp}.}
\bea
\left( \Delta_{E_{d+1}} -{6(4-d) (d+4)\over 8-d} \right)\, \cE_{(0,1)} (\mm)
= -\left ( \cE_{(0,0)} (\mm) \right )^2 + 40 \, \zeta(3)\,  \delta_{d,4} 
\label{eisenthree}
\eea
The quadratic term on the r.h.s. can be understood qualitatively \cite{Green:2005ba}
as a consequence of the $(\alpha')^3$  corrections to the supersymmetry variations,
although there has not been a precise derivation of this equation based on supersymmetry. 
The other terms 
on the second line are anomalous terms which arise for dimensions $d$ such that the 
eigenvalue of $ \cE_{(0,1)}$ vanishes or becomes degenerate with the eigenvalues 
of $ \cE_{(0,0)}$ or $ \cE_{(1,0)}$. They reflect the occurrence of logarithmic infrared 
divergences in string theory. The exact solutions to  \eqref{eisenthree} relevant for 
$\cE_{(0,1)}$ are not known explicitly, but the perturbative expansions are,  
as reviewed in the previous subsection.

\subsubsection{S-duality constraints on the perturbative coefficients}
\label{sec:sconstraints}

Having summarized the differential constraints satisfied by the exact coefficients $\cE_{(m,n)} (\mm) $ of
the $\cR^4$, $D^4\cR^4$ and $D^6\cR^4$ interactions in the low energy expansion,
it is now in principle straightforward to determine the differential constraints satisfied 
by the perturbative terms  $\cE^{(h)}_{(m,n)}(\rho_d)$ in the weak coupling expansion (\ref{2k4}).
For this purpose, however, it is important to include the contribution $\cE_{(m,n)}^{\rm non-an.}(g_d,\rr)$ of the terms proportional to powers of $\log \gd$. These terms can be computed
from the putative exact result, or from the non-local terms in the supergravity amplitude,
and are recorded below,\footnote{The coefficient of $\ln g_4$ in the last line agrees with the 
coefficient of the infrared singularity of the $D=6$ three-loop supergravity amplitude computed in \cite{Bern:2008pv}, thereby resolving a puzzle raised following equation (3.21)  in \cite{Green:2010sp}.}
\bea
\cE_{(0,0)}^{\rm non-an.} (\gd,\rr) &=& \frac{4\pi}{3}\, \ln g_2\, \delta_{d,2} 
\\
\cE_{(1,0)}^{\rm non-an.}  (\gd, \rr) &=&   \frac{16\pi^2}{15}\, \ln g_3\, \delta_{d,3} + \cE_{(0,0)}\, \ln g_4\, \delta_{d,4} 
\no \\
\cE_{(0,1)} ^{\rm non-an.} (\gd, \rr) &=& \left( - \frac{4\pi^2}{27} \ln^2 g_2 
+ \frac{2\pi}{9} \left( \frac{\pi}{2}+ \cE_{(0,0)} \right) \, \ln g_2 \right)\, \delta_{d,2}
 + 5 \, \zeta(3) \, \ln g_4\, \delta_{d,4}
 \qquad
\no
\eea
The action of the Laplacian $\Delta_{E_{d+1}}$ on $\cE_{(m,n)}$ can now be decomposed  
terms of the $SO(d,d)$ subgroup of $E_{d+1}$,
\be
\label{deltatoso}
\Delta_{E_{d+1}} = \frac{8-d}{8} \pa_\phi^2 + \frac{d^2-d+4}{4}\ \pa_\phi + \Delta_{SO(d,d)} + \cdots
\ee
where we have set $\gd = e^\phi$, and the ellipsis denotes derivatives with respect 
to the Ramond--Ramond moduli which 
are angular variables that decouple from perturbation theory. This agrees with 
\cite[A.24]{Obers:1999um}  for $d=2,3,4$,  upon  noting that for these values,
\be
2^{d-1} = \frac{2(d^2-d+4)}{8-d} 
\ee
Substituting (\ref{deltatoso}) into (\ref{eisenone})-(\ref{eisenthree}), we deduce 
that the $h$-loop contributions to  the  coefficients $\cE_{(m,m)}$ satisfy the following Laplace-type equations:
\begin{itemize}
\item
\noindent The perturbative parts of $\cE_{(0,0)} (\mm)$ satisfy
\bea
\label{delsor4}
\Delta_{SO(d,d)}  ~ \cE_{(0,0)} ^{(0)} (\rr) & = &  0
\\
\left( \Delta_{SO(d,d)}  + d(d-2)/2 \right)\, \cE_{(0,0)} ^{(1)} (\rr) & = & 4\pi\, \delta_{d,2} 
\no
\eea
\item
The perturbative parts of $\cE_{(1,0)} (\mm)$ satisfy 
\bea
\label{delsod4r4}
\Delta_{SO(d,d)}   ~ \cE_{(1,0)} ^{(0)} (\rr)  & = &  0 
\\ 
\left( \Delta_{SO(d,d)}  +(d+2)(d-4)/2 \right)\, \cE_{(1,0)} ^{(1)} (\rr) & = & 12\, \zeta(3)\, \delta_{d,4}
\no \\
\left( \Delta_{SO(d,d)}  + d(d-3) \right)\, \cE_{(1,0)} ^{(2)} (\rr) & = & 24\,\zeta(2)\, \delta_{d,3}
+ 4  \,\cE_{(0,0)} ^{(1)} (\rr) \, \delta_{d,4}
\no
\eea
\item
The perturbative parts of $\cE_{(0,1)} (\mm)$ satisfy 
\bea
\left( \Delta_{SO(d,d)}  -6 \right)\, \cE_{(0,1)} ^{(0)} (\rr)  &=&  -  \left ( \cE_{(0,0)} ^{(0)} (\rr) \right ) ^2
\\
\left( \Delta_{SO(d,d)}  - (d+4)(6-d)/2 \right)\, \cE_{(0,1)} ^{(1)} (\rr) &=&
-2 \cE_{(0,0)} ^{(0)} (\rr) \, \cE_{(0,0)} ^{(1)} (\rr) +\frac{2\pi}{3} \zeta(3)\, \delta_{d,2}
\nn\\
\left( \Delta_{SO(d,d)}  - (d+2)(5-d) \right)\, \cE_{(0,1)} ^{(2)} (\rr) &=&  - \left ( \cE_{(0,0)} ^{(1)} (\rr) \right ) ^2 -\left(\frac{\pi}{3} \cE_{(0,0)}^{(1)} + \frac{7\pi^2}{18} \right) \, \delta_{d,2}
\qquad
\nn\\
\left( \Delta_{SO(d,d)}  - 3 d(4-d)/2  \right)\, \cE_{(0,1)} ^{(3)} (\rr)  &=&  20 \zeta(3)\, \delta_{d,4} 
\label{delsod6r4}
\no \eea 
\end{itemize}
For all but the genus two $D^6 \cR^4$ amplitude $ \cE_{(0,1)} ^{(2)}(\rr) $ it is relatively 
straightforward to check that these equations are consistent with the Eisenstein 
series which describe the perturbative terms, as discussed in section~\ref{sec:pert}. 
In particular, the anomalous terms on the r.h.s. arise whenever the  Eisenstein series 
has a pole, after subtracting the contribution of the pole. In the case $d=0$, where 
the coefficients are constants and the Laplacian $\Delta_{SO(d,d)}$ vanishes, no such 
anomalous terms arise and  these equations 
reduce to algebraic relations between the coefficients.
In particular, the two-loop contribution  $\cE_{(0,1)} ^{(2)} (\rr)$ is predicted to take the value
stated in \eqref{twoloopcoeffs}.  

\sm

In the sequel we shall compute  the genus-two modular 
integral    (\ref{dsixgentwo}) for $d=0$ and check agreement with this prediction. The differential
equation on the third line of \eqref{delsod6r4} will indicate that the 
Zhang-Kawazumi invariant satisfies the differential equation \eqref{1c}, which holds
the key to the computation of the modular integral.

\section{The Zhang-Kawazumi invariant}
\label{zkprops}
\setcounter{equation}{0}

The ZK invariant $\f (\Sigma)$ may be defined on a surface $\Sigma$ of 
arbitrary genus $h \geq 2$ \cite{Zhang, Kawazumi}. 
As was already stated in the Introduction (\ref{defphi}), it  was shown in  \cite{D'Hoker:2013eea} 
to be given by the following expression $\f (\Sigma) = -  \int _{\Sigma ^2} P(x,y) \, G(x,y)/4h$. 
Here, the bi-form $P(x,y)$ is a section of $K\otimes \bar K$ in both $x$ and $y$, where $K$ 
is the canonical bundle on $\Sigma$,  and  may be expressed as,
\bea
\label{defP}
P(x,y) & = &  \sum _{I,J,K,L} P_{IJKL} \, \om _I(x) \, \overline{\om_J(x)} \, 
\om _K(y) \, \overline{\om_L(y)}
\no \\
P_{IJKL} & = & - Y_{IJ} ^{-1} \, Y^{-1}_{KL}  + h \, Y_{IL} ^{-1} \, Y^{-1}_{JK}
\eea
The  indices take the values $I,J,K,L=1,\cdots, h$; henceforth, repeated indices will be understood to be  summed. 
Furthermore, $\om _I$ is a basis of  canonically normalized holomorphic Abelian differentials, and $\Omega_{IJ}=X_{IJ}+i Y_{IJ}$
is the period matrix of the surface $\Sigma$ (see Appendix A for their detailed definitions.). The scalar Green function $G$ on $\Sigma ^2$ is given in terms of the prime form, $E(x,y)$, and the above quantities by,
\bea
\label{defG}
G(x,y) = - \ln |E(x,y)|^2 + 2 \pi Y^{-1}_{IJ} \left ( \Im \int _x ^y \om _I \right )
\left ( \Im \int _x ^y \om _J \right )
\eea
Its mixed derivatives are given by\footnote{Depending on context, we write $\om_I$ for the differential one form, 
or $\om_I(z)$ for the function representing the 1-form in a local coordinate system $(z,\bar z)$, 
so that the 1-form locally takes the form $\om _I = \om _I (z) dz$. In the latter case, the convention for the 
integral over $\Sigma$ includes a factor of the volume form $i dz \wedge d \bar z$, which will
not, however, be exhibited.}
\bea
\label{mixG}
\p_{\bar y} \p_x G(x,y) & = & + 2 \pi \delta (x,y) - \pi Y^{-1}_{IJ} \om _I(x) \, \overline{ \om _J (y)}
\no \\
\p_{\bar y} \p_y G(x,y) & = & - 2 \pi \delta (x,y) + \pi Y^{-1}_{IJ} \om _I(y) \, \overline{ \om _J (y)}
\eea
The bi-form  $P(x,y)$ is symmetric under the interchange of $x$ and $y$, and obeys the 
key property that its integral over a single copy of $\Sigma$ vanishes identically,
\bea
\label{3b2}
\int _{\Sigma _x} P(x,y)=0
\eea
The formula results from the combination of the following two elementary results,
\bea
\label{3b3}
\sum _{KL} P_{IJKL} \, Y_{KL}=0\,, \hskip 1in 
\int _\Sigma \om _K \wedge \overline{\om_L} = - 2 i Y_{KL}
\eea
The Green function is single-valued but transforms under conformal transformations
in $x$ by a shift which depends only upon $x$. Combining this transformation with
the property of (\ref{3b2}) guarantees that $\f (\Omega)$ in (\ref{defphi}) is well-defined.
In fact, in view of (\ref{3b2}), any properly normalized scalar Green function may be 
used instead of $G$, including the properly normalized Arakelov Green function 
(see \cite{D'Hoker:2013eea} for the detailed relations). 
Finally, we shall often write $\f (\Omega)$ instead of $\f (\Sigma)$ when we parametrize 
$\Sigma$ by its period matrix, by a slight abuse of notation.

\subsection{Differential constraint on the ZK invariant: a first hint}
\label{systemzk}

In order to gain insight into the nature of the Zhang-Kawazumi invariant $\f$, it is 
very instructive to examine the structure of  the differential equation satisfied by the 
genus-two coefficient  $\cE_{(0,1)}^{(2)}$ defined by the modular integral (\ref{dsixgentwo}). 
On the one hand, this coefficient  satisfies the equation in the third line of (\ref{delsod6r42}),  
which we reproduce below  for   $d\not=2$,
\be
\label{delsod6r42}
\left( \Delta_{SO(d,d)}  - (d+2)(5-d) \right)\, \cE_{(0,1)} ^{(2)} (\rr) =  - \left ( \cE_{(0,0)} ^{(1)} (\rr) \right ) ^2
\ee
On the other hand, the  lattice partition function for the torus $\TT^d$
satisfies the following differential equation in arbitrary genus $h$ \cite{Obers:1999um}, 
\be
\label{DeltaG}
\left ( \Delta_{SO(d,d)}  -2 \Delta_{Sp(2h)} + \half d h(d-h-1)  \right )\, \Gamma_{d,d,h} (\rr ; \Omega) = 0
\ee
where $\Delta_{Sp(2h)}$ is the Laplacian on the Siegel upper-half space $\cS_h$, 
defined in Appendix \ref{appA}.  Specializing to genus-two, where the moduli space $\cM_2$
coincides with  the Siegel upper half space $\cS_2$, and the Laplacian $\Delta$  
coincides with $\Delta _{Sp(4)}$, it follows from  (\ref{dsixgentwo}) and (\ref{DeltaG}) that
\be
\left( \Delta_{SO(d,d)}  - {(d+2)(5-d)} \right)\, \cE_{(0,1)} ^{(2)} (\rr) = 2
\pi \int_{\cM_2}  d \mu_2\, \varphi( \Omega)\, \left( \Delta - 5\right)\, 
\Gamma_{d,d,2} (\rr, \Omega)
\ee
Comparing with the differential equation (\ref{delsod6r42}), we see that agreement for $d\not=2$ requires, 
\be
\label{lapphi}
 \int_{\cM_2}  d \mu_2\, \varphi(\Omega)\,\left ( \Delta - 5\right )  
 \Gamma_{d,d,2}(\rr, \Omega) 
= -\frac{\pi}{2}\, \left( \int_{\cM_1}\, d \mu_1\, \Gamma_{d,d,1} (\rr, \tau) \right)^2 
\ee
After integration by parts, this becomes,
\bea
\label{lapphi2}
&&\int_{\cM_2} d \mu_2\, \Gamma_{d,d,2} (\rr, \Omega) \,\left ( \Delta - 5 \right )   \varphi(\Omega)
+ \int_{\partial\cM_2} \left(  \varphi\, \star d \Gamma_{d,d,2} - \Gamma_{d,d,2}  \star d  \varphi \right)
\nn\\
&&\hskip 6cm= -\frac{\pi}{2}\, \left( \int_{\cM_1}\, d \mu_1\, \Gamma_{d,d,1} (\rr, \tau) \right)^2 
\eea
The structure of this equation is very informative.  Recall that the boundary of $\cM_2$ 
includes the separating degeneration limit, $\Omega_{12}\to 0$, where  
$\Gamma_{d,d,2}(\rr; \Omega)\sim  \Gamma_{d,d,1}(\rr; \Omega_{11})\, \Gamma_{d,d,1}(\rr; \Omega_{22})$ has the factorised form of the right-hand side of the equation.
This is the first indication that the combination $\left ( \Delta - 5\right ) \varphi$ has 
support on boundary $\p \cM_2$ of moduli space.

\subsection{The ZK invariant in the supergravity limit}

Further evidence in support of the eigenvalue equation $(\Delta -5) \, \f=0$ in the interior of $\cM_2$
may be gathered by considering the limit of degenerating Riemann surfaces $\Sigma$. The 
Deligne-Mumford compactification of $\cM_2$ requires the addition to $\cM_2$ of just two 
divisors, namely the separating and the non-separating nodes. These nodes intersect,
and contain further degeneration divisors. One might study the fate of the equation $(\Delta -5) \, \f=0$
on any of these degenerations.

\sm

Here, we shall limit attention to studying the complete non-separating degeneration in which the components of $Y=\Im \, \Omega$ all become large, and the surface degenerates to two connected long thin tubes. This degeneration is physically significant, since it is directly related to the integrand of two-loop Feynman diagrams in $D=10$ supergravity. In particular the four-graviton amplitude in maximally supersymmetric theories can be rewritten in terms of graphs with cubic vertices. Ignoring the position of the external gravitons one obtains a skeleton graph with two tri-valent vertices. The lengths of the corresponding lines of the graph will be denoted by $L_i \gg 1$, and may be identified  with the entries of $Y$ as follows,
\be
\label{eq:omegaL}
 \Omega = i  Y_L  + \cO (1) 
 \hskip 1in 
 Y_L = \pmatrix{ L_1+L_3 & L_3 \cr L_3 & L_2+L_3 \cr} 
\ee
The $\cO(1)$ corrections which are being omitted here contain both the real parts of $\Omega$
as well as higher order corrections to $\Im \Omega$.  
To leading order in $L_i \gg 1$, the complete degeneration limit of the integral~(\ref{defphi}) may be 
expressed in terms of the graph lengths $L_i$ and the positions, $L_x$ and $L_y$, of the 
points $x,y$ which enter into the integral. The limiting behavior of Abelian differentials 
and of the prime form are standard, and have been discussed with the  help of the 
Schottky parametrisation in~\cite{Frizzo:1999zx}, as well as in the context of 
tropical modular geometry in~\cite{Tourkine:2013rda}. For our purposes it is sufficient to know the limits of the prime form and the Abelian differentials: $E(x,y)$ tends to the distance on the graph between the two insertion points multiplied (in our conventions) by $2\pi$, while ${\rm Im}\, \omega_1(x) = dL_x$ (respectively ${\rm Im}\, \omega_2(x) = dL_x$) if $x$ is on the thin tubes forming the first (respectively second) loop and zero otherwise.

\sm

The contribution to (\ref{defphi}) from the complete non-separating degeneration arises 
from two graph topologies: 
type (a) where the insertions $x,~y$ are on the same degenerating tube; and type (b) 
when they are on opposite tubes. The corresponding asymptotics of the Green function $G$
for both graphs were obtained in~\cite{Green:2008bf}, and are given by,\footnote{Here and below, 
the arrows encompass both taking the limit of large $L_i \gg 1$, as well as carrying out the angular 
integrations on moduli and the points $x,y$.}
\bea
\label{grefi}
G^{(a)}(x,y) & \to & 
G^{(a)}_L = -2\pi \left ( |L_x - L_y| - \frac{(L_2 + L_3) (L_x-L_y)^2}{\det Y_L} \right ) 
\\
G^{(b)}(x,y) & \to & 
G^{(b)}_L = -2\pi \left ( L_x + L_y - \frac{(L_1+L_3)L_y^2 + (L_2+L_3)L_x^2 + 2 L_x L_y L_3}{\det Y_L} \right ) 
\no
\eea
These formulas have been written down for type (a) when the points $x,y$ are on the tubes 
of length $L_1$; and for type (b) when $x$ is on the tube of length $L_1$ while $y$ is on 
the tube of length $L_2$. In the same limit the bi-form $P$ of~(\ref{defP}) on $\Sigma ^2$ 
for both types of graphs takes  the following form,
\bea
\label{Pfi}
P^{(a)}(x,y) &  \to & P^{(a)}_L dL_x dL_y =  - 4 \left(\frac{L_2+L_3}{\det Y_L}\right)^2 dL_x dL_y ~
\no \\
P^{(b)}(x,y) & \to &  P^{(b)}_L dL_x dL_y =  -4 \left(\frac{L_3^2-\det Y_L}{\det Y_L^2}\right) dL_x dL_y~
\eea
under the same assumptions on $x,y$ as we had spelled out for the Green function.

\sm

For both types of graph an overall factor of 2 arises from the possibility of exchanging the ordering of the two points $x,y$. 
Then there are other contributions with the same topology which are simply obtained by cyclical 
permutation of the $L_i$ and correspond to inserting the punctures on the other tubes forming the 
genus 2 degenerating surface. Thus from the diagrams of type (a) we have
\be
\varphi^{(a)}_L = -\frac{1}{4}
\int\limits_0^{L_1} \!\! dL_x \int\limits_0^{L_x} \!\! d L_y\, P^{(a)}_L G^{(a)}_L + {\rm cycl.} 
= 2\pi \left[-\frac{L_\Sigma}{12} +  \frac{L_1 L_2 L_3}{4 \det Y_L} +  \frac{L_1^2 L_2^2 L_3^2 L_\Sigma}{6 \det Y_L^3}\right]
\ee
where $L_\Sigma= L_1+L_2+L_3$. Similarly the contributions of the diagrams of type (b) is
\be
\varphi^{(b)}_L = -\frac{1}{4}
\int\limits_0^{L_1} \!\! dL_x \int\limits_0^{L_2} \!\! d L_y\, P^{(b)}_L G^{(b)}_L + {\rm cycl.} = 2\pi \left[\frac{L_\Sigma}{6} -  \frac{2 L_1 L_2 L_3}{ 3 \det Y_L} -  \frac{L_1^2 L_2^2 L_3^2 L_\Sigma}{6 \det Y_L^3}\right]
\ee
Adding up the two types of contributions $\f^{(a)}_L+\f^{(b)}_L=\f_L$ gives the following expression for $\f$ in the complete degeneration limit,
\bea
\label{kzmaxdeg}
\varphi(\Sigma) = \f _L + \cO(1) \hskip 0.7in \f_L=
\frac{\pi }{6}\, \left(L_1+L_2+L_3 - \frac{5\, L_1 L_2 L_3}{L_1 L_2+L_1 L_3+L_2L_3}\right) 
\eea
The contribution from the complete non-separating degeneration to the two-loop
$D^6 \cR^4$ effective interaction is therefore given by inserting the asymptotic expression 
(\ref{kzmaxdeg}) for $\f$ into the integral (\ref{dsixgentwo}), and setting the winding 
numbers $n^{\alpha I}$ in the lattice sum (\ref{Gamddh}) to zero.

\sm

The result should match the two-loop $D^6 \cR^4$ effective interaction in 10-dimensional 
supergravity compactified on $\TT^d$, which differs by a factor of $(\det Y)^{1/2}$ from 
the same interaction computed in 11-dimensional supergravity compactified on $\TT^d$.
The latter was computed in equation (3.6) of \cite{Green:2005ba} (or in equation 
(2.23) of \cite{Green:2008bf} after correcting a sign), and is in agreement with (\ref{kzmaxdeg}).
The asymptotic expression (\ref{kzmaxdeg}), in the limit where $L_2\gg L_1,L_3$, is also in agreement with the 
(double) degeneration limit  $\tau,\tau_1 \to i \infty$ of the non-separating degeneration formula (\ref{4a4}), which is already an expansion for $\tau_2 \to i \infty$.

\sm

To verify the equation $(\Delta - 5) \f=0$ in the complete degeneration limit,
it remains to evaluate the Laplacian $\Delta _L$ of (\ref{eq:Deltasph}) in this limit and we find, 
\be
\label{laplacedef}
\Delta_L = \sum_{i,j}L_iL_j  \frac{\partial}{\partial  L_i}\frac{\partial}{\partial  L_j}+\frac{ \det Y_L }{2}\, \left(\sum_{i=1}^3 \frac{\partial^2}{\partial  L_i^2} -2 \sum_{i<j}\frac{\partial}{\partial  L_i}\frac{\partial}{\partial  L_j} \right) 
\ee
It is easy to check that we indeed have $(\Delta_L - 5) \, \f_L=0$. 
With this additional encouragement we will now proceed to show 
that $\varphi$ does indeed satisfy (\ref{1c}) for all genus-two surfaces $\Sigma$.

\section{The Laplacian of the Zhang-Kawazumi invariant}
\setcounter{equation}{0}
\label{laplacezk}

The purpose of this section is to calculate, from first principles, the Laplacian of the 
ZK invariant $\f$, and show that $\f$ obeys the eigenvalue equation $(\Delta - 5) \, \f=0$ 
in the interior of the moduli space of genus-two Riemann surfaces $\cM_2$. 
The equation may be extended to the Deligne-Mumford compactification $\overline{\cM_2}$ 
at the cost of a right side which has support on the separating node. The derivation in this section
is somewhat technical and the hurried reader may wish to skip to section \ref{intvarphi}.

\subsection{Preliminaries}
\label{sec1}

The Laplacian of $\f$ will be computed using standard deformation theory of complex
structures on a Riemann surface. As it turns out, the problem may be formulated 
in arbitrary genus with little extra complication, and we shall carry out the calculations
in arbitrary genus. A brief summary of the Siegel
upper half space $\cS_h$ for arbitrary genus $h$, its Poincar\'e metric, volume form, 
action of the modular group, and sub-variety of the moduli space $\cM_h$ of 
compact genus $h$ Riemann surfaces is provided in Appendix \ref{appA}.

\sm

The $Sp(2h,\RR)$-invariant Laplace operator $\Delta _{Sp(2h)}$ on scalar functions on 
$\cS_h$ is defined in (\ref{eq:Deltasph}). For $h=2,3$ the $Sp(2h,\RR)$-invariance 
of the Laplacian $\Delta _{Sp(2h)}$ automatically induces a Laplacian on $\cM_2$ which 
we shall denote $\Delta$. For $h \geq 4$, the Laplacian $\Delta _{Sp(2h)}$ needs to be 
projected by restricting the derivatives $\p_{IJ}$ in (\ref{eq:Deltasph}) to the tangent space 
$T\cM_h$ at every point of $\cM_h$, and we shall denote the resulting Laplacian by $\Delta$.

\subsection{Basic variational formulas}

To evaluate the derivatives with respect to $\Omega$ and $\bar \Omega$
(projected onto $T \cM_h$ for $h \geq 4$), we shall use the standard theory and 
formulas of deformations
of complex structures. The tangent space $T\cM_h$ decomposes into a direct 
sum of holomorphic and anti-holomorphic subspaces, which are generated 
respectively by a Beltrami differential $\mu$ and its complex conjugate $\bar \mu$.  
Here, $\mu$ is a section of $K^{-1} \otimes \bar K$ where $K$ is the canonical bundle 
on $\Sigma$. To evaluate the Laplacian $\Delta $ on $\f$, we shall need to compute
the mixed variational derivatives of $\varphi$ with respect to $\mu$ and $\bar \mu$,
which automatically includes the needed projection from $T\cS_h$ to $T \cM_h$.

\sm

A holomorphic deformation $\delta _\mu \phi$  with Beltrami differential 
$\mu = \mu_{\bar w} {}^w d\bar w/dw $ of any function $\phi$ on $\cM_h$ is given as follows,
\bea
\delta _\mu \phi= { 1 \over 2 \pi} \int _\Sigma d^2w \,  \mu_{\bar w} {}^w  \, \delta _{ww} \phi
\eea
The deformations $\delta _{ww} \phi$ supported at the point $w$ may be viewed as resulting 
from the insertion of the stress tensor at the point $w$, and the particular normalization used
here is in accord with the standard normalizations of the stress tensor \cite{D'Hoker:1988ta}.

\sm

The point-wise deformation of the period matrix, the canonically normalized 
holomorphic Abelian differentials $\omega _I(x)$, and the prime form $E(x,y)$
are given as follows \cite{Verlinde:1986kw},
\bea
\label{var}
\delta _{ww} \omega _I (x)  & = & \omega _I(w) \p_x \p_w \ln E(x,w)
\no \\
\delta _{ww} \Omega _{IJ} & = & 2 \pi i \omega _I(w) \omega _J (w)
\no \\
\delta _{ww} \ln E(x,y) & = & - \half \Big ( \p_w \ln E(w,x) - \p_w \ln E(w,y) \Big )^2
\eea
The deformation of other quantities, such as Abelian differentials of the second
and third kind, may be obtained from the last equation by taking derivatives in $x$ and $y$.

\subsection{Calculation of the first variational derivative}

From the basic point-wise variational formulas of (\ref{var}), we now produce further 
variational formulas which will be of more direct utility in evaluating the Laplacian of $\f$.
We shall prefer to express the resulting formulas in terms of the single-valued Green
function $G$ rather than in terms of the multiple-valued prime form, and Abelian integrals.
First, one derives the following variational formulas for $G$ and $P$, 
\bea
\label{var0}
\delta _{ww} G(x,y) & = & \half \Big ( \p_w G(w,x) - \p_w G(w,y) \Big )^2
\no \\
\delta _{ww} P(x,y) & = & 
- P_{IJKL}  \om _I (w) \, \overline{\om_J(x)} \,  \om _K(y) \, \overline{\om_L(y)} \, \p_x \p_w G(w,x)
\no \\ &&
- P_{IJKL}  \om _I (x) \, \overline{\om_J(x)} \,  \om _K(w) \, \overline{\om_L(y)} \, \p_y \p_w G(w,y)
\eea
where $P_{IJKL} $ is the modular tensor defined in (\ref{defP}). A useful intermediate formula
in the derivation of both formulas in (\ref{var0})  is given by the relation, 
\bea
\delta _{ww} \left ( Y^{-1} _{IJ} \, \om _J (x) \right ) = - Y^{-1} _{IJ} \,
\om _J (w) \, \p_x \p_w G(w,x)
\eea
With the help of these
formulas, we obtain the first order variation of $\f$, 
\bea
\label{var1}
4 h \, \delta _{ww} \f & = & \int _{\Sigma ^2} \bigg \{
P_{IJKL} \, \om _I (x) \, \overline{\om_J(x)} \,  \om _K(y) \, \overline{\om_L(y)} \, \p_w G(w,x) \p_w G(w,y)
\no \\ && \hskip 0.3in 
- P_{IJKL} \, \om _I (w) \, \overline{\om_J(x)} \,  \om _K(y) \, \overline{\om_L(y)} \, \p_w G(w,x) \p_x G(x,y)
\no \\ && \hskip 0.3in 
- P_{IJKL} \, \om _I (x) \, \overline{\om_J(x)} \,  \om _K(w) \, \overline{\om_L(y)} \, \p_w G(w,y) \p_y G(x,y)
\bigg \}
\eea
Note that the terms proportional to $(\p_w G(w,x))^2$ and $(\p_w G(w,y))^2$, which
arise from the variation of $G$ in the first line of (\ref{var0}),  cancel in view of (\ref{3b2}).
Upon interchange of $x$ and $y$, the last two terms above are seen to be equal;
we have refrained from carrying through the corresponding simplification in order to 
retain the manifest symmetry 
under interchange of $x$ and $y$. With the help of the mixed derivative formulas for $G$
in (\ref{mixG}), one readily verifies holomorphicity of the first variation,  namely
$\p_{\bar w} \left ( \delta _{ww} \f \right ) =0$.

\subsection{Calculation of the second variational derivative}

To compute the mixed variation  $\delta _{\bar u \bar u} \delta _{ww} \f$,
we use the complex conjugated relations of (\ref{var}), but we also need to vary the 
derivatives with respect to the holomorphic coordinates. The starting point to do so
are the standard variational formulas for the Cauchy-Riemann operators $\p_z ^{(n)}$
and $\p_{\bar z} ^{(n)}$ on sections of $K^n$, which are given by \cite{D'Hoker:1988ta},
\bea
\delta _{\bar \mu} \p_{\bar z} ^{(n)} = 0
\hskip 1in 
\delta _{\bar \mu} \p_z ^{(n)} = \bar \mu \, \p_{\bar z} ^{(n)} + n \, ( \p_{\bar z} \bar \mu )
\eea
Here, we set $n=0$, drop the superscript $(n)$, and derive the point-wise deformations,
\bea
\delta _{\bar u \bar u} \p_{\bar z} = 0
\hskip 1in
\delta _{\bar u \bar u} \p_z = 2 \pi \, \delta (z,u) \p_{\bar z}
\eea
The key ingredients needed for the variation of (\ref{var1}) include the variation of a single 
derivative of $G$, which is found to be,
\bea
\label{derG}
\delta _{\bar u \bar u} \Big ( \p_x G(x,y) \Big ) = - \pi \p_{\bar x} \delta (u,x) 
+ \pi Y^{-1} _{IJ} \om_I(x) \, \overline{ \om _J(u)} \Big ( \p_{\bar u} G(u,y) - \p_{\bar u} G(u,x) \Big )
\eea
and the variation of the generalization of $P(x,y)$ which appears in (\ref{var1}), and  
which takes the following form,
\bea
\label{derP}
&&
\delta _{\bar u \bar u } 
\left ( P_{IJKL} \, \om _I (s) \, \overline{\om_J(x)} \,  \om _K(t) \, \overline{\om_L(y)}  \right )
\no \\ && \hskip 0.8in = 
- P_{IJKL}  \om _I (s) \, \overline{\om_J(u)} \,  \om _K(t) \, \overline{\om_L(y)} \, \p_{\bar u} \p_{\bar x} G(u,x)
\no \\ && \hskip 0.95in 
- P_{IJKL}  \om _I (s) \, \overline{\om_J(x)} \,  \om _K(t) \, \overline{\om_L(u)} \, \p_{\bar u}  \p_{\bar y} G(u,y)
\eea
for arbitrary points $s,t,x,y \in \Sigma$.

\sm

The calculation of these variational derivatives is fairly lengthy, and is relegated to 
Appendix \ref{appB}. The final result, valid for arbitrary genus, may be cast in the following form, 
\bea
\delta_{\bar u \bar u} \delta _{ww} \f  = \psi _A + \psi _B + \psi _C
\eea
where each one of these contributions is given by,
\bea
\label{4k1}
\psi _A & = & - { 2\pi \over 4h} (2h+2) \int _\Sigma \p_{\bar u} G(u,x) \, \p_w G(w,x)
\no \\ && \hskip 0.5in \times 
\left (  Y^{-1}_{IJ} \, Y^{-1}_{KL} - Y^{-1}_{IL} \, Y^{-1}_{JK}  \right )
\om _I (x) \, \overline{\om_J(x)} \,  \om _K(w) \, \overline{\om_L(u)} \,
\\
\psi _B & = &  {  \pi ^2 \over 2} \int _{\Sigma ^2} 
G(x,y) \, Y^{-1} _{CD} \, \om _C (x) \, \overline{ \om _D (u) } \,
Y^{-1} _{AB} \, \om _A (w) \, \overline{ \om _B (y) } 
\no \\ && \hskip 0.4in \times 
\Big (  Y^{-1} _{IJ} \, Y^{-1}_{KL} -  Y^{-1}_{IL} \, Y^{-1} _{JK}   \Big ) 
\om _I (w)  \, \om_K(y) \, \overline{ \om _J(x) } \,  \overline{ \om _L (u) }
\no \\
\psi _C & = & { 2 \pi \over 4 h} \int _{\Sigma ^2} \p_{\bar u} G(u,x) \, \p_w  G(w,y) \, P_{IJKL} Y^{-1}_{AB} 
\om _I(x) \overline{ \om _L (y)} 
\no \\ && \hskip 0.5in \times
\bigg \{ \om _K(w) \, \om_A(y) - \om _K(y) \, \om _A(w) \bigg \} 
\bigg \{ \overline{\om_J(u) } \, \overline{\om_B(x) } - \overline{\om_J(x) } \, \overline{\om_B(u) } \bigg \}
\no
\eea
On general grounds, the mixed derivative $\delta_{\bar u \bar u} \delta _{ww} \f $ 
satisfies the following three conditions,
\begin{enumerate}
\itemsep =-0.05in
\item Hermiticity, namely invariance under $w \leftrightarrow \bar u$;
\item Holomorphicity in $w$, namely $\p_{\bar w} (\delta _{\bar u \bar u} \delta _{ww} \f )=0$;
\item Holomorphicity in $\bar u$, namely $\p_u (\delta_{\bar u \bar u} \delta _{ww} \f )=0$.
\end{enumerate}
Hermiticity is seen to hold for each contribution $\psi_A, \psi_B, \psi _C$ separately.
Property 3 then follows from property 2, which in turn is proven in Appendix \ref{appB}.
Note that the holomorphicity properties are manifest for $\psi_B$, while in $\psi_A$ and 
$\psi_C$ they follow (at least in part) in view of the fact that the poles in the derivatives
$\p_w G$ and $\p_{\bar u} G$ are cancelled by manifest zeros of corresponding combinations
of Abelian differentials.

\subsection{Calculation of mixed variations for genus two}

For genus two, we may use the special properties of $h=2$ to further simplify the 
expressions for $\psi_A, \psi _B, \psi_C$. We shall make use of the bi-form $\Delta(x,y)$ 
which is defined by,
\bea
\label{3g1}
\om _I(x) \, \om_J(y) - \om _J(x) \, \om _I(y) = \ep _{IJ} \, \Delta (x,y)
\eea
Here, we have $I,J=1,2$, and we use the convention $\ep _{12}=1$. The bi-form 
$\Delta$ is a holomorphic section of the canonical bundle $K$ in both $x$ and $y$, and by construction its
existence is limited to genus-two. Its zeros are at $x=y$ and $x=I(y)$ where $I(y)$ is 
the image of $y$ under the hyper-elliptic involution of the genus-two surface $\Sigma$.
We shall also use the following relation, which is again special to genus-two, 
\bea
\label{3g2}
Y^{-1}_{IJ} \, Y^{-1}_{KL} - Y^{-1}_{IL} \, Y^{-1}_{JK} = \ep _{IK} \, \ep _{JL} \,  (\det Y)^{-1}
\eea
With the help of these relations we derive the following simplified expressions  for $h=2$,
\bea
\label{3g3}
\psi _A & = & -{3 \pi \over 2} (\det Y)^{-1} \int _\Sigma \p_{\bar u} G(u,x) \p_w G(w,x)
\Delta (x,w) \, \overline { \Delta (x,u)} 
\no \\
\psi _B & = & - {\pi^2 \over 2} (\det Y)^{-1} \int _{\Sigma ^2} G(x,y) Y^{-1} _{AB} \, Y^{-1}_{CD}
\, \om _A (w) \, \overline{\om_B(y)} \,  \om _C(x) \, \overline{\om_D(u)} 
\Delta (y,w) \, \overline { \Delta (x,u)} 
\no \\
\psi _C & = & +{3 \pi \over 4} (\det Y)^{-1} \int _{\Sigma^2} \p_{\bar u} G(u,x) \p_w G(w,y)\, 
\Delta (x,w) \, \overline { \Delta (x,u)} \, Y^{-1}_{IJ} \, \om_I(x) \, \overline{\om_J(y)} 
\eea
The next key observation is that, using the first line of (\ref{mixG}), 
we may combine  the first and the last lines of (\ref{3g3})  as follows,
\bea
\psi _A + \psi _C = -{3 \over 4} (\det Y)^{-1} \int _{\Sigma^2} \p_{\bar u} G(u,x) \p_w G(w,y)
\Delta (y,w) \, \overline { \Delta (x,u)} \p_x \p_{\bar y} G(x,y)
\eea
Upon integrating by parts in both $x$ and $\bar y$, and using the holomorphicity 
of $\Delta (y,w)$ in $y$ and of $\Delta (x,u)$ in $x$, we obtain an intgeral
involving the product $\p_x \p_{\bar u} G(u,x) \, \p_{\bar y} \p_w G(w,y)$. 
Using again (\ref{mixG}) on both mixed derivative factors, and exploiting the fact that the 
$\delta$-function contributions vanish, we find, 
\bea
\psi _A + \psi _C = -{3 \pi^2 \over 4} { Y^{-1} _{AB} \, Y^{-1}_{CD} \over \det Y} \int _{\Sigma^2} G(x,y)
 \, \om _A (w) \, \overline{\om_B(y)} \,  \om _C(x) \, \overline{\om_D(u)} 
\Delta (y,w) \, \overline { \Delta (x,u)}  
\eea
We recognize that this expression is proportional to $\psi_B$, so that our final formula
becomes, 
\bea
\delta_{\bar u \bar u} \delta _{ww} \f = -{5 \pi^2 \over 4} { Y^{-1} _{AB} \, Y^{-1}_{CD} \over \det Y} 
\int _{\Sigma^2} G(x,y)
 \, \om _A (w) \, \overline{\om_B(y)} \,  \om _C(x) \, \overline{\om_D(u)} 
\Delta (y,w) \, \overline { \Delta (x,u)}  
\eea
In this form, hermiticity and holomorphicity in $w$ and $\bar u$ are manifest properties.

\subsection{Calculation of $\Delta \f$ for genus two}
\label{sec:32}

The form $\delta_{\bar u \bar u} \delta _{ww} \f $ is a holomorphic quadratic differential in $w$ and in $\bar u$. 
For genus-two, a basis of holomorphic quadratic differentials may be chosen in terms of the Abelian 
differentials, namely $\om _I (w) \om _J(w)$ in $w$ for $I\leq J$, and similarly in $\bar u$. 
To exhibit this dependence systematically, we introduce the following notation, 
\bea
\delta_{\bar u \bar u} \delta _{ww} \f  & = &
4 \pi^2 \om _I(w) \, \om _J (w) \, \overline{ \om _K (u) } \, \overline{ \om _L (u) }\,
T_{IJ;KL| AB;CD} \, \Phi_{AB;CD}
\no \\
\Phi _{AB;CD} & = & - { 5 \over 64} 
\int _{\Sigma ^2} G(x,y) \, \om_A(x) \, \overline{ \om_B (x)} \, \om_C(y) \, \overline{ \om_D (y)} 
\eea
The tensor $T$ is defined as follows for all genera,
\bea
\label{Tensor}
T_{IJ;KL| AB;CD} & = & 
+  Y^{-1} _{ID} \, Y^{-1}_{KA}  \Big ( Y^{-1} _{JL} \, Y^{-1}_{BC} - Y^{-1} _{JB} \, Y^{-1}_{LC} \Big )
\no \\ && 
+ 
 Y^{-1} _{JD} \, Y^{-1}_{KA}  \Big ( Y^{-1} _{IL} \, Y^{-1}_{BC} - Y^{-1} _{IB} \, Y^{-1}_{LC} \Big )
\no \\ && 
+
 Y^{-1} _{ID} \, Y^{-1}_{LA}  \Big ( Y^{-1} _{JK} \, Y^{-1}_{BC} - Y^{-1} _{JB} \, Y^{-1}_{KC} \Big )
\no \\ && 
+
 Y^{-1} _{JD} \, Y^{-1}_{LA}  \Big ( Y^{-1} _{IK} \, Y^{-1}_{BC} - Y^{-1} _{IB} \, Y^{-1}_{KC} \Big )
\eea
Note that the four terms in $T$ arise from the symmetrization conditions
in $I,J$ and $K,L$.

\sm

From the above formula, and the variational formula for the period matrix in (\ref{var}),
we deduce the partial derivatives of $\f$ with respect to $\Omega$ and its complex conjugate,
\bea
\bar \p _{KL} \p_{IJ} \varphi =  T_{IJ;KL| AB;CD} \, \Phi _{AB;CD}
\eea
The Laplacian, defined in (\ref{eq:Deltasph}), may now be applied to $\f$, and we find,
\bea
\Delta \f =  4\, Y_{IK}\, Y_{JL} T_{IJ;KL| AB;CD} \, \Phi _{AB;CD}
\eea
The contraction of the tensors yields the tensor $P_{ABCD}$ defined in (\ref{defP}), 
\bea
\label{YYT}
Y_{IK}\, Y_{JL} T_{IJ;KL| AB;CD} = 2 P_{ABCD}
\eea
Putting all together, and using 
the definition and normalization of $\f$ in (\ref{defphi}), we derive the Laplace eigenvalue 
equation for $\f$ that we had set out to prove, 
\bea
\label{4f1}
\Delta \, \f = 5 \, \f
\eea
We note that for genus higher than 2, no such simple expression appears to
be available. In Appendix \ref{appC}, we push the calculation of the corresponding
equation for genus $h \geq 3$ as far as possible. From a purely string theory point of view, 
of course, the ZK invariant is a natural object for genus-two, but probably not for higher genus.

\section{Integrating the ZK invariant over moduli space}
\setcounter{equation}{0}
\label{intvarphi}

The purpose of this section is to provide a first principles calculation of the integral
of the ZK invariant $\f$ over the moduli space $\cM_2$ of genus-two compact Riemann surfaces, 
$\int _{\cM_2} d \mu_2 \, \f$, 
and thus to prove directly from superstring perturbation theory the value  for this integral in 
(\ref{predicttwo}) and (\ref{1b}) predicted from the interplay between S-duality and supersymmetry.

\sm

The key new ingredient we shall use here is the Laplace eigenvalue equation (\ref{4f1})
satisfied by $\f$ in the interior of $\cM_2$. This equation allows us to recast the integral 
of $\f$ over $\cM_2$ in terms of an integral of $\Delta \f$ over $\cM_2$, and this last
integral can be reduced to an integral over the boundary $\p \cM_2$ of moduli space.
We now proceed to do so, properly taking into account convergence issues and
contributions from the boundary $\p \cM_2$.

\subsection{Convergence and regularization near the separating node}

The integral $\int _{\cM_2} d \mu_2 \, \f$ is absolutely convergent, a property established 
already in \cite{D'Hoker:2013eea}. Convergence may be  verified explicitly by recalling
the behavior of the volume form $d \mu_2$ and the ZK invariant $\f$ near the separating 
and non-separating components of the Deligne-Mumford compactification divisor of $\cM_2$.
To this end, it will be convenient to parametrize the period matrix $\Omega$ and the 
volume form $d \mu_2$ as follows,
\bea
\label{4a1}
 \Omega =  \pmatrix{ \tau_1 & \tau\cr  \tau & \tau_2\cr}
 \hskip 1in
 d\mu_2 = { d^2 \tau \, d^2 \tau_1\,  d^2 \tau_2 \over (\det Y)^3}
\eea 
where $d^2 \tau = i d \tau \wedge d\bar \tau$ and so on.
To leading order, the asymptotics of the volume form $d \mu_2$ is governed by the following expansions,
\bea
\label{4a2}
\hbox{separating} \hskip 0.2in & \hskip 0.5in & \det Y = \Im (\tau_1) \Im (\tau_2) + \cO (\tau ^2)
\no \\
\hbox{non-separating} & \hskip 0.5in & \det Y = \Im (\tau_1) \Im (\tau_2) + \cO (\tau_2^0)
\eea
The asymptotics of $\f$ near the separating node is given by, 
\bea
\label{4a3}
\f (\Omega) = - \ln \Big | 2 \pi \tau \, \eta (\tau_1)^2 \eta (\tau_2)^2 \Big | + \cO \left ( |\tau |^2 \ln |\tau| \right )
\eea
while near the non-separating node $\f$ has the following asymptotics, derived in \cite{D'Hoker:2013eea} 
and \cite{DeJong1} with the help of the degeneration results of  \cite{Wentworth:1991},
\bea
\label{4a4}
\f (\Omega) & = & { \pi \over 6 } (\Im \tau_2) + { 5 \pi \over 6} { (\Im \tau)^2 \over (\Im \tau_1)}
- \ln \left | {\tet _1 (\tau, \tau_1) \over \eta ( \tau_1)} \right | + \cO \Big ( (\Im \tau_2)^{-1} \Big )
\eea
Using the parametrization in terms of $\tau, \tau_1, \tau_2$ of the fundamental domain 
for $\cM_2$ given in Appendix \ref{appA}, it follows by inspection that the integral 
$\int _{\cM_2} d \mu_2 \, \f$ is absolutely convergent. 

\sm

To circumvent having to deal with modifications supported on the boundary $\p \cM_2$
to the Laplace eigenvalue equation $(\Delta -5)\f=0$ , we shall work with a regularized
integral, which is kept away from the boundary. We shall prove below that no  contributions 
arise from the non-separating node, so we need to regularize only near the separating node. 
To this end, we introduce the regularized domain for  moduli space,  defined by,
\bea
\label{4a5}
\cM_2 ^\ep =  \cM_2 \cap  \Big \{\tau \in \CC, ~ |\tau|> \ep  \Big \}
\eea
Everywhere on the space $\cM_2^\ep$, 
the function $\f$ satisfies $\Delta \f - 5 \f =0$, just as it did in the interior of $\cM_2$.
Since the integral $\int _{\cM_2} d \mu_2 \, \f$ over all of moduli space is absolutely convergent, 
we may recast it as a limit as $\ep \to 0$ of integrals over $\cM_2^\ep$ instead, 
and for finite $\ep$ use the Laplace eigenvalue equation, 
\bea
\label{4a6}
\int _{\cM_2} d \mu_2 \, \f =  \lim _{\ep \to 0}  \int _{\cM_2^\ep } d \mu_2 \, \f 
= {1 \over 5} \, \lim _{\ep \to 0}  \int _{\cM_2^\ep } d \mu_2 \, \Delta \f 
\eea
This equation will be the starting point for reducing the integral of the ZK invariant
to an integral over the boundary of the regularized moduli space $\cM_2^\ep$.

\subsection{Reducing the integral to the boundary of moduli space}

To analyze the  contribution from the boundary of moduli space 
arising from $\Delta \f$, we use the following form of the  Laplacian acting on scalars,
\bea
\label{5b1}
(\det Y)^{-3} \Delta  = 2 \bar \p_{IJ} \Big ( (\det Y)^{-3} Y_{IK} Y_{JL}  \p _{KL} \Big )  + {\rm c.c}
\eea
The formula follows directly from the usual differential geometry expression for the 
Laplacian with metric $ds^2 = g_{\alpha \beta} dx^\alpha dx^\beta$ being given by
$\sqrt{g} \Delta =  \p_\alpha ( \sqrt{g} g^{\alpha \beta} \p_\beta )$ where $g=\det g_{\alpha \beta}$.
It may also be easily verified directly with the help of (\ref{eq:Deltasph}).
It will be convenient to recast the formula using the following notations, 
\bea
\label{5b2}
(\det Y)^{-3}\Delta =   2 \p_{\bar \tau_1}  \dd_{\tau_1} + 2 \p_{\bar \tau_2}  \dd_{\tau_2} 
+ 2\p_{\bar \tau}  \dd_\tau + {\rm c.c.}
\eea
where the first order differential operators $\dd$ are defined  by,
\bea
\label{5b3}
\dd_{\tau_1}  & = & (\det Y)^{-3} \, Y_{1K} Y_{1L} \, \p_{KL}
\no \\
\dd_{\tau_2}  & = & (\det Y)^{-3} \, Y_{2K} Y_{2L} \, \p_{KL}
\no \\
\dd_{\tau} ~ & = & (\det Y)^{-3} \, Y_{1K} Y_{2L} \, \p_{KL}
\eea
Fortunately, we may answer the issue of boundary contributions by using
only the leading asymptotic behaviour of $\Delta$. Sub-leading terms are typically 
difficult to compute. Keeping only the leading behaviour of the pre-factor $\det Y$, but 
exactly in all other contributions, we then have for both degenerations,
and in terms of the coordinates $\tau, \tau _1, \tau_2$, 
\bea
\label{5b4}
\dd_{\tau_1}  & = & (\Im \tau_1 )^{-3} (\Im \tau_2)^{-3} 
\Big ( (\Im \tau _1)^2  \p_{\tau _1} 
+ (\Im \tau _1) (\Im \tau) \p_{\tau}  
+ (\Im \tau)^2 \p_{\tau_2} \Big )
\no \\
\dd_{\tau_2}  & = & (\Im \tau_1 )^{-3} (\Im \tau_2)^{-3} 
\Big ( (\Im \tau _2)^2 \p_{\tau _2} 
+ (\Im \tau _2) (\Im \tau) \p_{\tau}  
+ (\Im \tau)^2 \p_{\tau_1} \Big )
\no \\
\dd_\tau & = & (\Im \tau_1 )^{-3} (\Im \tau_2)^{-3} 
\bigg ( \half \left \{ (\Im \tau _1) (\Im \tau_2) + (\Im \tau)^2 \right \}  \p_{\tau } 
\no \\ && \hskip 1.5in 
+ (\Im \tau _1) (\Im \tau) \p_{\tau_1}  
+ (\Im \tau_2) (\Im \tau) \p_{\tau_2} \bigg  )
\eea
We must now investigate the behavior of $\dd_\tau \f$  as $\tau \to 0$, 
while keeping $\tau_1,\tau_2$ fixed for the separating node,  
and the behavior of $\dd_{\tau_2} \f$  as $\tau_2 \to i \infty$, 
while keeping $\tau, \tau_1$ fixed for the non-separating node.
To do so, we use the asymptotic behaviors of (\ref{4a3}) and (\ref{4a4}).

\sm

Near the separating node, we find a pole as $\tau \to 0$, 
\bea
\label{5b5}
\dd_\tau \f = (\Im \tau_1)^{-2} (\Im \tau_2)^{-2} \left ( - { 1 \over 4 \tau}  + \cO (|\tau| \ln |\tau|) \right )
\eea
Near the non-separating node, we find a contribution that tends to zero as $\tau_2 \to i \infty$, 
\bea
\label{5b6}
\dd_{\tau_2} \f = - { i \pi \over 12} (\Im \tau_1)^{-3} (\Im \tau_2)^{-1} + \cO \Big ( (\Im \tau_2)^{-2} \Big )
\eea
We conclude from this that the non-separating degeneration node does not 
contribute to the integral of $\Delta \f$. On the other hand, however, there is a contribution 
from a pole at the separating degeneration node.

\subsection{Calculation of the integral $\int d\mu_2 \, \f$}

In the preceding sections, we have shown that the integral of (\ref{4a6}) 
receives contributions only from the pole that arises at the separating node,
while the contribution from the non-separating node vanishes identically. To extract the
contribution from the pole at the separating node, we make use of the boundary expression 
for the Laplacian,
\bea
(\det Y)^{-3} \Delta \f & \approx & 2 \p_{\bar \tau } \dd_\tau \f + 2 \p_\tau \dd_{\bar \tau} \f
\eea
or in terms of the measure,
\bea
d \mu_2 \,  \Delta \f & \approx & 2 d^2\tau \,  d^2\tau_1 \,  d^2\tau_2 \,  
\left ( \p_{\bar \tau } \dd_\tau \f +  \p_\tau \dd_{\bar \tau} \f \right )
\eea
where we use the convention $d^2 \tau = i d \tau \wedge d \bar \tau$.
By integration by parts in $\tau$ over $\cM_2^\ep$, the boundary contribution 
localizes at $|\tau|=\ep$. To extract it, we make use of the relation,
\bea 
\I \, d \tau \wedge d \bar \tau \left ( \p_{\bar \tau } \dd_\tau \f +  \p_\tau \dd_{\bar \tau} \f \right )
= d \left ( - \I \, d \tau \, \dd_\tau \f + \I \, d \bar \tau \dd_{\bar \tau} \f \right )
\eea
so that 
\bea
 \int _{\cM_2} d \mu_2 \, \f =  {2 \over 5} \,
\lim _{\ep \to 0}  \int _{\p \cM_2^\ep }  d^2 \tau_1 \,  d^2 \tau_2 \, 
\Big (  \I \, d \tau \, \dd_\tau \f - \I \, d \bar \tau \, \dd_{\bar \tau} \f \Big )
\eea
where $\p \cM_2^\ep$ stands for the boundary $|\tau|=\ep$ near the separating node,
and is given by,
\bea
\p \cM_2 ^\ep =  \{\tau \in \CC, ~ |\tau|=\ep\}  \times (\cM_1 \times \cM_1) / (\ZZ_2 \times \ZZ_2)
\eea
Using the above results for $\dd_\tau \f $, 
\bea
\dd_\tau \f = (\Im \tau_1) ^{-2} (\Im \tau_2)^{-2} v_\tau  \hskip 1in v_\tau = - { 1 \over 4 \tau}
\eea
and the value for the genus-one volume,
\bea
\int _{\cM_1} { |d \tau_i|^2 \over (\Im \tau_i)^2} = { 2 \pi \over 3}
\eea
for $i=1,2$, we may perform the integrations over $\tau_1$ and $\tau_2$ first, and we get,
\bea
 \int _{\cM_2} d \mu_2 \, \f = {2 \over 5} \times {1 \over 2^2} \times 
 \left ( { 2 \pi \over 3 } \right )^2 \lim _{\ep \to 0} 
\oint _{|\tau |=\ep} \Big (  \I \, d \tau \, v_\tau - \I \, d \bar \tau \, v_{\bar \tau} \Big )
\eea
The value of the integral on the rhs is simply $\pi$, so that we have,
\bea
\int _{\cM_2} d \mu_2 \, \f = { 2 \pi ^3 \over 45}
\eea
Upon multiplying both sides by a factor of $\pi$, we 
reproduce the value announced in (\ref{predicttwo}) and predicted in  \cite{D'Hoker:2013eea}.

\subsection{Differential relation for the Faltings invariant}

In this subsection, we shall provide immediate mathematical implications of the relations derived above.
First we extend the validity of the eigenvalue equation for $\f$ to the compactified moduli
space $\overline{\cM}_2$. Second, we derive Laplace eigenvalue equations  
for two other Siegel modular forms, including the Faltings invariant for genus-two surfaces.

\sm

To begin, we complete the analysis of the ZK invariant by quoting the Laplace eigenvalue equation
valid on the Deligne-Mumford compactification $\overline{\cM_2}$ of moduli space. 
Combining the result $(\Delta - 5 ) \f=0$ of (\ref{4f1}) with the observation that the contribution from
the non-separating node vanishes, while the contribution of the separating node results from 
combining equations (\ref{5b2}), and (\ref{5b5}), we find, 
\bea
\Delta \f - 5 \f = - 2 \pi \delta _{SN} 
\hskip 1in 
\delta _{SN} = (\det Y) \, \delta ^{(2)} (\tau)
\eea
Here the Dirac $\delta$-function has been normalized to $\int d^2\tau \, \delta ^{(2)}(\tau)=1$,
and $\delta _{SN}$ is the induced $\delta$-function on the separating node.

\sm

The relation of $\f (\Omega)$ with the Faltings invariant $\delta (\Omega) $ was established 
in \cite{DeJong2}, 
\bea
\f (\Omega) = 36 \ln 2 - 40 \ln (2 \pi) - 3 \ln \| \Psi _{10} (\Omega) \| - { 5 \over 2 } \delta (\Omega)
\eea
Here, $\Psi _{10}$ is the genus-two cusp form of Igusa, and its Peterson norm is defined by,
\bea
\| \Psi _{10} (\Omega) \| = (\det Y)^5 |\Psi _{10} (\Omega)|
\eea
The Laplacian of its logarithm is given by, 
\bea
\Delta \ln \| \Psi_{10} (\Omega) \| = -15 + 4 \pi \delta _{SN}
\eea
The $\delta_{SN}$-function comes from the separating node, where $\ln |\Psi _{10} (\Omega)|
\approx \ln |\tau|^2$.
An alternative expression for $\f$ was obtained in \cite{D'Hoker:2013eea}
in terms of an integral over the Jacobian of $\Sigma$, 
\bea
\f (\Omega) & = & 6 \ln 2 -{1 \over 4} \ln |\Psi _{10} (\Omega)|^2 + 5 \ln \Phi (\Omega)
\no \\
\ln \Phi (\Omega) & = & \int _{\TT^4} d^4 x \ln \Big | \tet [x](0,\Omega) \Big |^2
\eea
where the characteristics $x$ take values in the unit square torus $\TT^4$.

\sm

It is now immediate to obtain the corresponding relations for the Faltings invariant and for 
the modular form $\Phi$, and we have,
\bea
(\Delta - 5 ) \, \delta (\Omega) & = & \f_1 + 6 \ln \| \Psi _{10} (\Omega)\| - 4 \pi \delta _{SN}
\no \\
(\Delta -5) \, \ln \Phi (\Omega) & = & 6 \ln 2 - \half \ln |\Psi _{10} (\Omega)|
\eea
where $\f_1=18-72\ln 2 +80 \ln (2 \pi)$. Note that the equation for  $\Phi$ has vanishing 
contribution from the separating divisor, as is indeed consistent with its regularity at this 
node \cite{D'Hoker:2013eea}.

\vskip 0.3in

\noindent
{\bf \large Acknowledgments}

\bigskip

We would like to acknowledge useful conversations with Robin De Jong, Michael Gutperle, and Arnab Rudra.
One of us (ED) thanks the Kavli Institute for Theoretical Physics at the University of California, 
Santa Barbara for their hospitality and the Simons Foundation for their financial support while part
of this work was being carried out. The research of ED was supported in part by the National 
Science Foundation under grants PHY-13-13986 and PHY-11-25915.  MBG  acknowledges 
funding from the European Research Council under the European 
Community's Seventh Framework Programme (FP7/2007-2013) / ERC grant agreement no. [247252].
The research of RR has been partially supported by the Science and Technology Facilities 
Council Consolidated (Grant ST/J000469/1, String theory, gauge theory \&  duality).


\appendix

\section{Some modular geometry}
\setcounter{equation}{0}
\label{appA}

The Siegel upper half space $\cS_h$ of rank $h$ is defined by,
\bea
\label{3a1}
\cS_h = \{ \Omega _{IJ} = \Omega _{JI} , ~ 1\leq I,J \leq h, ~\quad \Im , \Omega >0 \}
\eea
Symplectic transformations  $M \in Sp(2h,\RR)$ are defined by the relations,
\bea
\label{3a2}
M^t J M=J \hskip 0.7in 
M = \left ( \matrix{ A & B \cr C & D } \right )
\hskip 0.7in 
J = \left ( \matrix{ 0 & I_h \cr -I_h & 0  } \right )
\eea
The action of $Sp(2h,\RR)$ on $\Omega$ is given by,
\bea
\Omega \to M(\Omega) = (A\Omega+B) (C \Omega +D)^{-1}
\eea
The isotropy group of any point in $\cS_h$ is isomorphic to $U(h)$, so that 
$\cS_h$ may also be viewed as a coset $\cS_h = Sp(2h,\RR)/U(h)$.
We decompose $\Omega$ into real matrices $X,Y$ via $\Omega = X + i Y$,
and to use the abbreviation $Y^{-1}_{IJ} = (Y^{-1})_{IJ}$. 

\subsection{Metric and volume}

The Poincar\'e metric on $\cS_h$ is constructed to be invariant under $Sp(2h,\RR)$, and is given by
\bea
\label{3a3}
ds_h^2 = \sum _{I,J,K,L=1}^h Y^{-1}_{IJ} \, d \bar \Omega _{JK} \, Y^{-1}_{KL} \, d \Omega _{LI}
\eea
The associated invariant volume form  $d \mu_h$ is defined by,
\bea
\label{A3}
d \mu_h = { 1 \over (\det Y)^{h+1} } \bigwedge _{I\leq J} i \, d \Omega _{IJ} \wedge d \bar \Omega _{IJ}
\eea
It may be readily verified, for example, that $ds_h^2$ and $d\mu_h$ are invariant under 
the scaling transformation $\Omega \to \lambda ^2 \Omega$, for which we have 
$A=D^{-1} = \lambda I_h$ and $B=C=0$. 

\sm

The Siegel fundamental domain $\cF_h$ is the quotient of the Siegel upper half space
$\cS_h$ by the modular group 
$Sp(2h,\ZZ)/\ZZ_2$. Its volume $V_h = \int _{\cF_h} d \mu_h $ was calculated by 
Siegel \cite{Siegel},  and is given by,\footnote{The normalization of  the volume form
in (\ref{A3})  includes an extra factor of $2$ for each complex moduli dimension,
as compared to the volume form used in \cite{Siegel} and  \cite{Angelantonj:2011br,Pioline:2014bra}.} 
\bea
\label{volumedefs}
V_h = 2 \prod _{k=1}^h \left (  {2^k \over \pi ^k} \, \Gamma (k) \, \zeta (2k) \right )
\eea
where $\zeta(2k)$ is the Riemann zeta function. Its lowest values for even argument are given by, 
\bea
\label{zeta}
\zeta(2)= {\pi^2 \over 6}  \hskip 0.8in 
\zeta (4) = {\pi^4 \over 90} \hskip 0.8in 
\zeta (6) = {\pi^6 \over 945}
\eea 
which results in the following values of the 
volumes, 
\bea
\label{vols}
V_1 = { 2 \pi \over 3} \hskip 0.8in V_2 = {4 \pi^3 \over 3^3 \, 5} \hskip 0.8in 
V_3 = {2^6 \pi^6 \over 3^6 \, 5^2 \, 7}
\eea

\subsection{The Laplace-Beltrami operator}

The $Sp(2h,\RR)$-invariant Laplacian on $\cS_h$,
which is associated with the Poincar\'e metric, was derived in \cite{Maass} and is given by,
\bea
\label{eq:Deltasph}
\Delta_{Sp(2h)} =  \sum_{I,J,K,L=1}^h 4\, Y_{IK} \, Y_{JL} \, \bar \p_{KL} \, \p_{IJ}
\eea
Throughout, we shall use the standard composite index notation $\p_{IJ}$ for the partial derivatives 
with respect to $\Omega$, defined by,
\bea
\p_{IJ} \equiv \half \left ( 1 + \delta _{IJ} \right ) { \p \over \p \Omega _{IJ} }
\eea
along with their complex conjugates $\bar \p_{IJ}$. The above definition of the 
derivative in $\Omega$ guarantees that the derivative behaves in a tensorial manner under 
modular transformations, so that for example we have 
\bea
\p_{IJ} \Omega _{KL} = \half \left ( \delta _{IK} \delta _{JL} + \delta _{IL} \delta _{JK} \right )
\eea
We note that the Laplacian is normalised so that, 
\be 
\label{eq:Deltanor}
\Delta_{Sp(2h)} \, (\det Y )^s = \frac12 h s (2s-h-1)\, (\det Y )^s
\ee
It readily follows that $ \Delta_{Sp(2h)} \ln (\det Y ) = -\half h(h+1)$.

\subsection{Moduli spaces of low genus}

We give a synopsis of the moduli spaces $\cM_h$ of compact Riemann surfaces $\Sigma$
at low genus~$h$ and their representation as fundamental domains $\cF_h$ in $\cS_h$.
This will allow us to easily identify their canonical metric, volume form, and volume. To define 
these ingredients, we fix a homology basis of 1-cycles $A_I, B_I$ with $I=1,\cdots, h$
in $H_1(\Sigma, \ZZ)$, 
with canonical intersection pairing $\#(A_I,A_J)=\#(B_I, B_J)=0$,  $\#(A_I, B_J) = \delta _{IJ}$,
and a dual basis of holomorphic 1-forms $\omega _I$ in $H^1(\Sigma, \ZZ)$ subject to the 
canonical normalization on $A_I$-cycles, 
\bea
\oint _{A_I} \omega _J = \delta _{IJ} 
\hskip 1in 
\oint _{B_I} \omega _J = \Omega _{IJ}
\eea
The period matrix $\Omega$ belongs to $\cS_h$. For a given Riemann surface, the 
period matrix is defined up to a modular transformation $M \in Sp(2h,\ZZ) \subset Sp(2h,\RR)$, 
which corresponds to a redefinition of the canonical homology basis in $H_1(\Sigma, \ZZ)$,
with integer coefficients. Note that the element $-I \in Sp(2h,\ZZ)$ leaves every $\Omega$
invariant, so the identification is more properly under $Sp(2h,\ZZ)/\ZZ_2$.

\sm

The moduli space $\cM_1$ of genus-one compact Riemann surfaces 
coincides with the fundamental domain of $Sp(2,\ZZ)/\ZZ_2$ in $\cS_1$ and is given explicitly by,
\bea
\cM_1 = \cF_1=
\left \{ \tau \in \bC; ~ \Im (\tau) >0, ~ |\tau|\geq 1, ~ |\Re (\tau)|\leq \half \right \}
\eea
The $Sp(2,\RR)$-invariant Poincar\'e metric $ds_1^2$ is that of (\ref{3a3}) for genus-one, 
the associated volume form $d \mu_1$ is that of (\ref{A3}) for genus one, and total volume is 
$V_1$, as given in (\ref{vols}).

\sm

The moduli space $\cM_2$ for compact genus-two Riemann  surfaces  may be identified 
with the fundamental domain $\cF_2$ of $Sp(4,\ZZ)/\ZZ_2$ in $\cS_2$. Actually, the separating 
node $\cM_1 \times \cM_1$ must be removed, as it does not correspond to a compact surface. 
$\cM_2$ may be described concretely by the following set of inequalities on $\Omega = X+iY$, 
which were established in \cite{Gottschling1959} and were reviewed in \cite{Klingen},
\bea
\label{domain}
(1) &&  -\half \leq X_{11}, X_{12}, X_{22} \leq + \half
\no \\
(2) && 0 <  2 Y_{12}  \leq Y_{11} \leq Y_{22}
\no \\ 
(3) &&  |\det ( C \Omega +D) | \geq 1 ~ \hbox{for all} ~ \left ( \matrix{
A & B \cr C & D \cr } \right ) \in Sp(4,\ZZ)
\eea
The $Sp(4,\RR)$-invariant Poincar\'e metric $ds_2^2$ is that of (\ref{3a3}) for genus-two,
and the associated volume form $d \mu_2$ is that of (\ref{vols}) for genus-two. The 
volume $V_2$ is given in (\ref{vols}).

\sm

The moduli space $\cM_3$ for compact genus-three Riemann surfaces may be 
identified with the fundamental domain of $Sp(6,\ZZ)/\ZZ_2$ in $\cS_3$.
Actually, this identification is two-to-one, and requires the removal of the 
some sub-varieties. As a result, the volume is $V_3/2$ where $V_3$ was given in (\ref{vols}).
For higher genus, $h \geq 4$, the dimensions of $\cM_h$ 
and $\cS_h$ are respectively $3h-3$ and $h(h+1)/2$ and no longer match. Instead, 
$\cM_h$ is then a complex sub-variety of $\cS_h$ specified by the Schottky relations.

\section{Calculation of the mixed variation of $\f$}
\setcounter{equation}{0}
\label{appB}

In this Appendix, we provide the details of the calculation of the mixed variational
derivatives $\delta _{\bar u \bar u} \delta _{ww} \f$, and prove that this form is 
holomorphic in $w$ and $\bar u$. The starting point is the expression for the 
first variational derivative of $\f$ given in (\ref{var1}), along with the variational
derivatives given in (\ref{derG}) and (\ref{derP}).

\sm

The contributions to $\p_{\bar u \bar u} \delta _{ww} \f$ may be split
into those arising from the variations of the terms of the form 
$P_{IJKL} \, \om _I (s) \, \overline{\om_J(x)} \,  \om _K(t) \, \overline{\om_L(y)}$ in (\ref{var1}),
and those arising from the variation of the derivatives of $G$ in (\ref{var1}).
In an obvious notations, we have,
\bea
\label{B1}
\delta_{\bar u \bar u} \delta _{ww} \f = \delta_{\bar u \bar u} \delta _{ww} \f \Big |_P 
+ \delta_{\bar u \bar u} \delta _{ww} \f \Big |_G
\eea
These contributions may be written explicitly. For the part involving $P$, we have,
\bea
\label{var2}
\delta_{\bar u \bar u} \delta _{ww} \f \Big |_P 
& = & { 1 \over 2 h} \int _{\Sigma ^2} P_{IJKL} \bigg \{
\\ && \hskip 0.2in 
- \, \om _I (x) \, \overline{\om_J(u)} \,  \om _K(y) \, \overline{\om_L(y)} \,
\p_w G(w,x) \, \p_w G(w,y) \, \p_{\bar u} \p_{\bar x} G(u,x) 
\no \\ && \hskip 0.2in
+ \, \om _I (w) \, \overline{\om_J(u)} \,  \om _K(y) \, \overline{\om_L(y)} \,
\p_w G(w,x) \, \p_x G(x,y) \, \p_{\bar u}  \p_{\bar x} G(u,x) 
\no \\ && \hskip 0.2in
+ \, \om _I (x) \, \overline{\om_J(u)} \,  \om _K(w) \, \overline{\om_L(y)} \,
\p_w G(w,y) \, \p_y G(x,y) \, \p_{\bar u}  \p_{\bar x} G(u,x) \bigg \}
\no
\eea
Here we have used the symmetry under the interchange between the integrations
over $x$ and $y$, and the indices  $P_{IJKL}=P_{KLIJ}$ to pairwise combine terms
and bring out an  overall factor of 2.
For the part involving the variations of the derivatives of $G$ in (\ref{var1}), 
one first establishes that the contributions from the terms in $\bar \p \delta$ in (\ref{derG}) 
vanish identically. The remaining part simplifies to give the following result, 
\bea
\label{var3}
\delta_{\bar u \bar u} \delta _{ww} \f \Big |_G 
& = & {  \pi \over 2 h} \int _{\Sigma ^2} P_{IJKL} \, Y^{-1} _{AB} \, \bigg \{
\\ && \hskip 0.3in 
+ \om _I (x) \, \overline{\om_J(x)} \,  \om _K(y) \, \overline{\om_L(y)} \,  \om _A(w) \, \overline{\om_B(u)} \,
\p_{\bar u} G(u,x) \, \p_w G(w,y)
\no \\ && \hskip 0.3in 
+ \om _I (w) \, \overline{\om_J(x)} \,  \om _K(y) \, \overline{\om_L(y)} \,  \om _A(w) \, \overline{\om_B(u)} \,
G(x,y) \, \p_{\bar u}  \p_x G(u,x)
\no \\ && \hskip 0.3in 
- \om _I (w) \, \overline{\om_J(x)} \,  \om _K(y) \, \overline{\om_L(y)} \,  \om _A(x) \, \overline{\om_B(u)} \,
\p_{\bar u} G(u,y) \, \p_w G(w,x) \bigg \}
\no 
\eea
To obtain the cancellation of the $\bar \p \delta$ terms, and some further simplifications, we
have used the fact that $G$ is single-valued, so that integrations by parts can be performed
without producing boundary terms (since $\Sigma$ has no boundary), as well as the 
orthogonality relations of (\ref{3b3}).  Further simplifications are obtained as follows. 
In (\ref{var2}), we integrate by parts in $\bar x$ to eliminate one of the Green function
contributions using (\ref{mixG}). The mixed derivatives that
arise in the process produce $\delta$-functions which, using (\ref{B6}) below, yield $\psi _A$. 
They also produce terms that are of the same form as the terms in (\ref{var3}), and which are 
combined in the terms $\psi _B$ and $\psi _C$ below, as follows,
\bea
\label{B2}
\delta_{\bar u \bar u} \delta _{ww} \f  = \psi _A + \psi _B + \psi _C
\eea
The expressions for $\psi_A$ and $\psi _C$ are those given in the main body of the paper
(\ref{4k1}). For $\psi_B$, we find, 
\bea
\label{B3}
\psi _B & = & { 2 \pi \over 4 h} \int _{\Sigma ^2} G(x,y) \, \p_x \p_{\bar u} G(u,x) \, P_{IJKL} \, Y^{-1}_{AB} \,
\om _I(w) \, \overline{ \om _L (y)} 
\no \\ && \hskip 0.5in \times 
\om_K(y)  \, \om _A(w) 
\bigg \{ \overline{\om_J(x) } \, \overline{\om_B(u) } - \overline{\om_J(u) } \, \overline{\om_B(x) } \bigg \}
\eea
The single poles in the derivatives of the Green functions at $x=u,w$ in $\psi _A$
and at $\bar x = \bar u$ and $y=w$ in  $\psi_C$ are cancelled by the zeros of the Abelian differential
factors due to the antisymmetry under the interchange of $I,K$ and independently
of $J,L$. The $\delta (x,u)$-function arising from the mixed derivative of $G$ in $\psi_B$
is similarly cancelled by the effect of antisymmetry in $J,B$.
The last simplification leads to the expression for  $\psi _B$ in (\ref{4k1}).

\sm

The mixed derivative $\delta_{\bar u \bar u} \delta _{ww} \f $ must 
satisfy hermiticity, namely invariance under $w \leftrightarrow \bar u$, and 
holomorphicity in $w$ and $\bar u$. Hermiticity is established by inspection of each 
contribution $\psi_A, \psi_B, \psi _C$ separately.
Holomorphicity in $w$ is manifest for the contribution $\psi_B$ in (\ref{4k1}), as its only $w$-dependence
is through the holomorphic Abelian differentials $\om_A(w) \om _I(w)$. 
The other contributions are readily evaluated using the second equation in (\ref{mixG}),
\bea
\label{B5}
\p_{\bar w} \psi _A &=& 
 - { 2\pi^2 \over 4h} (2h+2) Y^{-1}_{CD} \, \om_C(w) \, \overline{\om _D (w)} \int _\Sigma \p_{\bar u} G(u,x) \, 
\\ && \hskip 0.5in \times 
\left (  Y^{-1}_{IJ} \, Y^{-1}_{KL} - Y^{-1}_{IL} \, Y^{-1}_{JK}  \right )
\om _I (x) \, \overline{\om_J(x)} \,  \om _K(w) \, \overline{\om_L(u)} \,
\no \\
\p_{\bar w} \psi _C & = & {  \pi^2 \over 2 h} Y^{-1}_{CD} \, \om_C(w) \, \overline{\om _D (w)} \, 
\int _{\Sigma ^2} \p_{\bar u} G(u,x) \,   P_{IJKL} Y^{-1}_{AB}  \, \om _I(x) \, \overline{ \om _L (y)} 
\no \\ && \hskip 0.5in \times
\bigg \{ \om _K(w) \, \om_A(y) - \om _K(y) \, \om _A(w) \bigg \} 
\bigg \{ \overline{\om_J(x) } \, \overline{\om_B(u) } - \overline{\om_J(u) } \, \overline{\om_B(x) } \bigg \}
\no
\eea
Carrying out the integration over $y$ in $\p_{\bar w } \psi _C$, and using (\ref{3b3}) and  the identity
\bea
\label{B6}
P_{IJKB} - P_{IBKJ} = - (h+1) \Big ( Y^{-1}_{IJ} \, Y^{-1}_{KB} - Y^{-1}_{IB} \, Y^{-1}_{JK}  \Big )
\eea
it is immediate that $\p_{\bar w} \psi _C = - \p_{\bar w} \psi _A$, thereby proving holomorphicity in $w$.

\section{Calculation of the Laplacian of $\f$ for genus $h \geq 3$}
\setcounter{equation}{0}
\label{appC}

Mathematically, the ZK invariant may be defined in terms of the scalar Green function
for any $h \geq 3$. Physically, however, there is no compelling evidence at this time that 
the ZK invariant plays any role in superstring perturbation theory at genus $h \geq 3$. 
Even if it did,  it is not even clear to which order $p$ in $D^{2p}\cR^4$ it would contribute.
Still, from our variational approach, we have access to evaluating the Laplacian on moduli
space $\cM_h$ of $\f$, and we shall carry as far as possible its calculation in this Appendix.

\sm

To evaluate $\Delta \f$ for all genera, we shall derive a formula which isolates the 
part that contributes for $h \geq 3$, but vanishes identically for $h=2$. It is in this
form that the higher genus expression for $\Delta \f$ will remain as close as possible
to the genus-two formula. To do so, we transform $\psi_A$ in (\ref{3g3})  into a double 
integral over $\Sigma ^2$, just as $\psi _B, \psi _C$ are double intgerals, by using the 
second identity in (\ref{mixG}).
The resulting two contributions of $\psi_A$ will be denoted respectively by $\psi _A ^1, \psi _A^2$ with, 
\bea
\label{C2}
\psi _A = \psi _A^1 + \psi _A^2
\eea
The double integrals are given as follows,
\bea
\label{C3}
\psi _A^1 & = & 
- { 2h+2 \over 4h}  \int _{\Sigma^2}  \p_{\bar u} G(u,x) \, \p_w G(w,x) \p_x \p_{\bar y} G(x,y)
\no \\ && \hskip 1in \times 
\left (  Y^{-1}_{IJ} \, Y^{-1}_{KL} - Y^{-1}_{IL} \, Y^{-1}_{JK}  \right )
\om _I (y) \, \overline{\om_J(x)} \,  \om _K(w) \, \overline{\om_L(u)} 
\no \\
\psi _A^2 & = & 
- { \pi \over 4h} (2h+2) \int _{\Sigma^2} \p_{\bar u} G(u,x) \, \p_w G(w,x) \, Y^{-1}_{AB} \om_A(x) 
\overline{\om_B(y)} 
\no \\ && \hskip 1in \times 
\left (  Y^{-1}_{IJ} \, Y^{-1}_{KL} - Y^{-1}_{IL} \, Y^{-1}_{JK}  \right )
\om _I (y) \, \overline{\om_J(x)} \,  \om _K(w) \, \overline{\om_L(u)} 
\eea
Integrating by parts in $x$ and $\bar y$ in $\psi _A^1$, we recover a term 
proportional to $\psi _B$. With the help of the modular tensor $T$, originally defined in 
(\ref{Tensor}) for genus-two but now extended to all genera, and after some further simplifications, we obtain,
\bea
\label{C4}
\psi _A^1 + \psi _B & & =
- \pi ^2 \, { 2h+1\over 8h} \,
\om_I(w) \, \om_J(w) \, \overline{ \om_K(u)} \, \overline{ \om_L(u)} \, T_{IJKL| ABCD} 
\no \\ && \hskip 1in \times
\int _{\Sigma ^2} G(x,y) \, \om_A (x) \, \overline{ \om_B(x)} \, \om_C(y) \, \overline{ \om_D(y)} 
\eea
We extract the part of the Laplacian due to $\psi _A^1+\psi_B$, 
in a manner which generalizes the calculation
of section \ref{sec:32} to arbitrary genus. In particular, we make use of the contraction
formula (\ref{YYT}) which was established for genus-two, but in fact holds unmodified for 
arbitrary genus. The result is as follows,
\bea
\label{C5}
\Delta \f \bigg | _{\psi _A^1 + \psi _B} = (2h+1) \f
\eea
Note that this expression already saturates the equation for $h=2$, so we should
expect the contribution from $\psi _A^2 +\psi _C$ to the Laplacian to vanish for $h=2$.

\subsection{The contribution to $\Delta \f$ which vanishes for $h=2$}

The remaining part $\psi = \psi _A^2 + \psi _C$ of $\delta _{\bar u \bar u} \delta _{ww} \f$ 
may be combined as follows,
\bea
\psi  =
{ 2 \pi \over h} S_{IJK;ABC}
\int _{\Sigma ^2} \p_{\bar u} G(u,x) \, \p_w  G(w,y) \,  
\om _I(x) \, \om _{[J} (w) \, \om _{K]}(y) \,
\overline{\om_{[A} (u) } \, \overline{\om_{B]} (x) } \, \overline{ \om _C (y)}
\eea
where the brackets $[{ ~} ]$ instruct to  anti-symmetrize the enclosed indices,
namely  $JK$ in the first brackets, and $AB$ in the second.
The modular tensor $S$ arising from combining equations (\ref{3g3}) for $\psi_C$
and (\ref{C3}) for $\psi _A^2$ is given by the anti-symmetrization in the indices
$AB$ and $JK$ of the expression, 
\bea
- 4 Y^{-1}_{IA} \, Y^{-1}_{JC} \, Y^{-1}_{KB}  - 2 (h-1) Y^{-1}_{IC} \, Y^{-1}_{JB} \, Y^{-1}_{KA} 
\eea
Remarkably, the properly anti-symmetrized form may be simply expressed in terms 
of the unique rank-six, degree three anti-symmetric tensor of $Y^{-1}$, defined by,
\bea
\mA _{IJK;ABC} & \equiv & +
Y^{-1} _{IA} \, Y^{-1} _{JB}  \, Y^{-1} _{KC} + 
Y^{-1} _{IB} \, Y^{-1} _{JC}  \, Y^{-1} _{KA} + 
Y^{-1} _{IC} \, Y^{-1} _{JA}  \, Y^{-1} _{KB} 
\no \\ &&
- Y^{-1} _{IA} \, Y^{-1} _{JC}  \, Y^{-1} _{KB}  
- Y^{-1} _{IB} \, Y^{-1} _{JA}  \, Y^{-1} _{KC}  
- Y^{-1} _{IC} \, Y^{-1} _{JB}  \, Y^{-1} _{KA} 
\eea
One finds, 
\bea
S_{IJK;ABC} =  \mA _{IJK;ABC} +  Y^{-1} _{IC} \, Y^{LD} \, \mA_{JKL;ABD}
\eea
For $h=2$, the tensor $\mA_{IJK;ABC}$ vanishes identically, so that the contribution
the $\Delta \f$ originating from $\psi$ vanishes identically as well. To check that
$\psi $ is holomorphic in $w$ and $\bar u$, it suffices to make use of the identity,
\bea
Y^{IB} \, S_{IJK;ABC} = 0
\eea
and its permutations.
The part of $\Delta \f$ coming from $\psi$ may be evaluated by using the holomorphicity
of $\psi$ in $w$ and $\bar u$ to write $\psi$ as follows,
\bea
\psi = 4 \pi^2 \, \om_I (w) \, \om _J (w) \, \overline{\om _K (u)} \, \overline{ \om _L(u)} \, 
\Lambda _{IJ; KL}
\eea
so that  we obtain our final formula, 
\bea
\Delta \f - (2h+1) \f = 4 Y_{IK} \, Y_{JL} \, \Lambda _{IJ; KL}
\eea
To extract $\Lambda _{IJ;KL}$ from $\psi$ is cumbersome, but straightforward.
Using the fact that $\psi $ is a holomorphic quadratic form in both $w$ and $\bar u$,
and that the set $\om _I (w) \om _J(w)$ and 
$\overline{\om _K (u)} \, \overline{ \om _L(u)}$ for $I,J,K,L=1,\cdots, h$ spans a basis for 
such forms, we may obtain the $w$-dependence by choosing $3h-3$ generic points
$p_a$ at which to evaluate $w$, and $3h-3$ generic points $q_b$ at which to evaluate $u$.
$\Lambda _{IJ;KL}$ is then obtained by inverting the $(3h-3) \times (3h-3)$-dimensional 
matrix. The procedure will provide a unique modular tensor $\Lambda_{IJ;KL}$ for $h=3$,
but will fail to give a unique result for $h \geq 4$ due to the Schottky relations,
which may be viewed as imposing $\half (h-2)(h-3)$ linear relations between the 
quadratic differentials $\om _I (w) \om _J(w)$.

\newpage


\end{document}